\documentclass[article]{aa}                           

\usepackage[varg]{txfonts}
\usepackage{graphicx}
\usepackage{array}
\usepackage{lscape}                                
\usepackage{txfonts}
\usepackage{caption}

\usepackage{longtable}
\usepackage{color}
\usepackage{supertabular}
\usepackage[draft]{hyperref}

\usepackage[flushleft]{threeparttable}
\usepackage{multirow}
\usepackage{lscape}
\usepackage{comment}
\usepackage{booktabs}
\usepackage[normalem]{ulem}

\newcommand{\Mv} {$M_{\rm V}$}
\newcommand{\Av} {$A_{\rm V}$}
\newcommand{\BCv} {$BC_{\rm V}$}
\newcommand{\Msp} {$M_{\rm sp}$}
\newcommand{\kms} {km\,s$^{-1}$}

\newcommand{\vsini} {$v$\,sin\,$i$}

\newcommand{\Teff} {$T_{\rm eff}$}
\newcommand{\grav} {log\,{\em $g$}}
\newcommand{\gravt} {log\,{\em $g_{\rm true}$}}

\newcommand{\fastwind} {{\sc fastwind}}
\newcommand{\ioni}[2]{{#1\,\sc{#2}}}

\newcommand{\msol}{$M_{\odot}$}
\newcommand{\Lsp} {log\,($\mathcal{L}$/$\mathcal{L_{\odot}}$)}
\newcommand{\Llum} {log\,(${L}$/${L_{\odot}}$)}

\newcommand{\numO}{{358}}
\newcommand{\numCalib}{{234}}
\newcommand{\numElim}{{124}}

\begin{document}
\title{The IACOB project}

\subtitle{XV. Updated calibrations of fundamental parameters of Galactic O-type stars
} 

\author{G.~Holgado\inst{1,2}, S.~Sim\'on-D\'iaz\inst{1,2}, A.~Herrero\inst{1,2}}

\institute{
            Instituto de Astrof\'isica de Canarias, E-38200 La Laguna, Tenerife, Spain.
             \and
             Departamento de Astrof\'isica, Universidad de La Laguna, E-38205 La Laguna, Tenerife, Spain.
             }
		   
\offprints{gholgado@iac.es}

\date{Date}

\titlerunning{Fundamental parameters of Galactic O-type stars}
\authorrunning{Holgado et al.}

\abstract
{O-type stars are key drivers of galactic evolution. For the first time, modern spectroscopic surveys combined with \textit{Gaia} distances are enabling reliable estimates of the fundamental parameters of several hundred Galactic O-type stars, spanning the full range of spectral types and luminosity classes.}
{We aim to provide updated, statistically robust empirical calibrations of the fundamental parameters of Galactic O-type stars, as well as of their absolute visual magnitudes ($M_V$) and bolometric corrections, based on high-quality observational data.}
{We performed a homogeneous analysis of a sample of 358 Galactic O-type stars, combining high-resolution spectroscopy and \textit{Gaia} distances.
A subset of 234 stars meeting strict quality criteria involving parallax, extinction, and multiband photometry was used to derive empirical calibrations of fundamental parameters with respect to spectral type (SpT) and luminosity class (LC).
We also evaluated the utility of the FW3414 parameter (based on the shape of the H$\beta$ line profile) as a calibrator for $M_V$, useful in large surveys lacking reliable spectral classification. 
As a consistency test, we compared radii, luminosities, and spectroscopic masses derived from two independent \Mv\ calibrations, one based on spectral classification and the other on FW3414, with their directly determined counterparts.
}
{We present updated 
SpT-based calibrations of fundamental parameters for 
LCs V, III, and I. 
Compared to previous works, we find systematic shifts, particularly in effective temperature for dwarfs and in $M_V$ across all classes.
Significant scatter in \Mv\ persists even with Gaia DR3 distances, which propagate into derived quantities. 
Applying the SpT\,--\,\Mv\ and FW3414\,--\,\Mv\ calibrations to the full sample yields consistent estimates of radius and luminosity, while the spectroscopic mass ($M_{\rm sp}$) shows significant scatter. 
}
{These updated empirical calibrations offer a robust reference for Galactic O-type stars and will support studies of massive star populations in both Galactic and extragalactic contexts, particularly in the era of large spectroscopic surveys.}


\maketitle

\section{Introduction}\label{section1}

O-type stars populate the upper main-sequence region of the Hertzsprung-Russell diagram \citep{Humphreys1978} and are key agents of galactic evolution. Their intense radiation fields, strong winds, and short lifetimes shape the interstellar medium and regulate star formation. Accurate knowledge of their fundamental parameters is essential not only for stellar evolution and feedback studies, but also for interpreting observations of massive star populations in both Galactic and extragalactic environments.

Initially thought to be just normal hydrogen-burning stars born with masses above $\sim$\,15\,\msol, recent studies suggest that a modest fraction may also result from complex scenarios, such as the merger of two less massive stars or from stars that become gainers in binary systems following a mass transfer event \citep{deMink2014}. O-type stars are also frequently found in single (SB1) and double-line spectroscopic binary (SB2) systems, or as part of hierarchical multiple systems \cite[e.g.,][]{Mason2009, Chini2012, Sana2013, Sana2014, Aldoreta2015, Barba2017, Sana2017, MaizApellaniz2019}. 

Although scarce when compared to their lower-mass counterparts \citep{Salpeter1955}, they are key contributors to the ionization of the surrounding interstellar medium \citep{Stromgren1939}. This is a consequence of their high effective temperatures (\Teff), which range from approximately 30\,000 to 55\,000 K. 
Given their short lifetimes, on the order of a few million years, these stars are predominantly located in regions of active star formation, including both young clusters and OB associations \citep{Blaauw1964, Wright2023, Quintana2024}. However, a significant fraction are also observed as runaway stars, having migrated from their birthplaces \citep{Blaauw1961, Stone1991, GiesBolton1986, MaizApellaniz2018, CarreteroCastrillo2023}.

Since the advent of stellar atmosphere models \citep[see][for a review]{Mihalas2001}, several studies have strived to establish simple physical calibrations for O-type stars, linking fundamental parameters, such as effective temperature, surface gravity, luminosity, radius, and stellar mass, with their spectral morphology. A key motivation is to provide reliable methods of obtaining initial estimates of the stellar properties for newly discovered O-type stars \citep[e.g.,][]{Doran2013, Shenar2024} based on their readily accessible spectral type (SpT) and luminosity class (LC). This approach avoids the need to perform complex and often time-consuming quantitative spectroscopic analyses \citep{SimonDiaz2020}, while also overcoming the usual situation in which reliable information about distances is not accessible.

Among the most widely cited works on Galactic O-type stars are those by \cite{Conti1973}, \cite{Panagia1973}, \cite{Vacca1996}, and \citet{Martins2005}. The latter has become a standard reference following the development of modern stellar atmosphere codes for massive blue stars \citep[see reviews by][]{Sander2017, PulsHerreroAllende2024}. Additionally, significant progress has been made in calibrating the SpT\,–-\,\Teff\ relation for O-type stars in various Local Group galaxies, with metallicities ranging from approximately half to one-tenth solar \citep[e.g.,][]{Massey2005, Heap2006, Mokiem2007, Massey2009, Garcia2010, RiveroGonzalez2012, Camacho2016, Sabin-Sanjulian2017, Ramirez-Agudelo2017,Lorenzo2025}. 
Two significant caveats have limited progress in refining these calibrations. The first is the poor statistical robustness of the samples used, which complicates attempts to accurately estimate the intrinsic scatter naturally expected in such calibrations \citep{SimonDiaz2014a}. The second, particularly relevant for stars in the Milky Way until recently (see below), has been the lack of reliable distance measurements for large stellar samples. This limitation has hindered precise determination of their luminosities, radii, and masses, with published (averaged) uncertainties on the order of  
0.25 dex, 15\%, and 40\%,
respectively \citep[e.g.,][]{Repolust2004,Marcolino2009,Bouret2012,Mahy2015,Markova2018}. 
Additionally, an important question arises, of how the substantial fraction of binary interaction products expected to populate the O-star domain affects the uniqueness and reliability of these calibrations. 

Thanks to significant efforts led by the IACOB project \citep{SimonDiaz2020a}, the number of Galactic O-type stars with spectroscopically derived parameters, namely, \Teff\ and surface gravities, \grav, has increased by approximately an order of magnitude over the last decade, reaching nearly 400 homogeneously analyzed targets \citep{Holgado2018, Holgado2020}. Furthermore, IACOB, in collaboration with complementary observational projects such as GOSC and GOSSS \citep{MaizApellaniz2011, MaizApellaniz2013}, OWN \citep{Barba2017}, and MONOS \citep{MaizApellaniz2019}, and supported by numerous other studies in the literature, has facilitated an improved classification of the investigated sample into presumably single stars and spectroscopic binaries (both SB1 and SB2).

\begin{figure}[!t]
\includegraphics[width=0.5\textwidth]{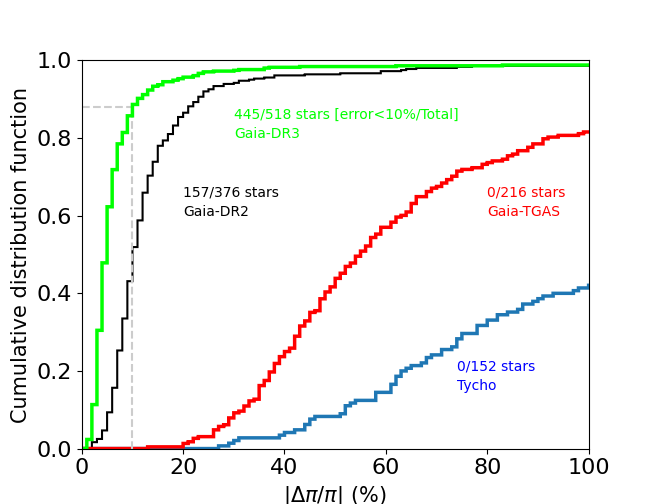}
\caption{Cumulative histograms of the nominal relative error in parallax of increasing large samples of Galactic O-type stars with available parallaxes following the Tycho-2 \citep{Hog2000} astrometric catalogs, and the various \textit{Gaia} data releases, including the \textit{Gaia}-TGAS solution \citep{Michalik2015}. For each catalog we indicate the total number of known stars of this type with available entries, along with those in which a relative error in parallax $<$10\% was achieved (see also vertical line).}
\label{Paralaje_Gaia}
\end{figure}

The {\em Gaia} mission \citep{Gaia2016,Gaia2018,Gaia2021,Gaia2023} has also marked a major milestone in providing reliable parallaxes for Galactic O-type stars, especially since the second data release (DR2). As is illustrated in Fig.~\ref{Paralaje_Gaia}, prior to {\em Gaia}-DR2, hardly any of the $\approx$\,600 known O-type stars in the Milky Way \citep{MaizApellaniz2013, MaizApellaniz2016} had parallax measurements with relative errors $<$40\%. This situation has improved drastically, with a significant fraction of these stars now benefiting from distance estimates with accuracies better than 
5\,--\,10\%.

The time is ripe to revisit the physical calibrations for Galactic O-type stars proposed by \citet[][hereafter M05]{Martins2005}. In \citet{Holgado2018}, we presented an initial revision of their SpT\,--\,\Teff\ and \grav\ relations, based on the sample of standard stars for spectral classification, last updated by \cite{MaizApellaniz2016} and comprising a total of 128 stars. In this paper, we expand the analysis to include the full sample of O-type stars previously studied within the framework of the IACOB project. 
Additionally, we leverage a curated list of reliable \textit{Gaia} distances to refine all previously established calibrations, including the absolute visual magnitude (\Mv) scale and other fundamental parameters such as radii, luminosities, and spectroscopic masses.

Dedicated studies focusing on individual clusters -- such as the recent work by \citet{Berlanas2025} on Carina~OB1 using \textit{Gaia}~DR3 parallaxes -- are essential to assess and validate distance-dependent methodologies. While such works offer high precision in specific environments, our broader Galactic-scale approach is better suited for deriving general empirical calibrations applicable to the full O-star population.

The paper is structured as follows. 
In Sect.~\ref{section2}, we describe the calibrator sample used in this study.
Sect.~\ref{section3} outlines the methodology applied to derive fundamental stellar parameters from the available data.
Sect.~\ref{section4} presents the resulting calibrations, including their validation and a discussion on the effects of binarity.
Section~\ref{section5} provides our conclusions and some future prospects.

\section{Sample description}\label{section2}

As of now, the IACOB database contains $\sim$17000 high-resolution spectra for 532 of the 621 O-type stars listed in version 4.1 of the Galactic O-Star Catalog (GOSC; \citealt{MaizApellaniz2013})
The working sample we initially considered for the study presented in this paper comprises those \numO\ Galactic O-type stars among the 532 which have at least one high-resolution spectrum available in the IACOB spectroscopic database \citep{SimonDiaz2020, Holgado2020}, and fulfill two additional criteria; namely, that they are not identified as clear double-lined (or higher-order) spectroscopic systems, and they are not considered peculiar or extreme systems. 
In particular, the latter criterion refers to those stars with peculiarities in their spectra \cite[classified as, e.g., Oe, Ope, Ofp, or Of?p,][]{Sota2011, Sota2014, MaizApellaniz2016} as well as those with SpTs earlier than O4. Those were excluded from the sample because their high effective temperatures compromise the diagnostic lines considered by our analysis strategy to determine spectroscopic parameters \citep[see][]{Holgado2020}.

Among the 358 stars, 82 were identified as SB1 systems based on radial velocity variations detected across all available spectra\footnote{Within the sample of 358 stars, for $\sim$75\% of them we have at least three spectra and for $\sim$15\%, only one. See the specific number of available spectra per star in Table~\ref{tableValues}.}, while accounting for the possibility that some variability could originate from intrinsic stellar phenomena \citep{SimonDiaz2024}. Further details about the strategy we followed to obtain estimates of the radial velocity per available spectra and identify clear SB1 systems can be found in \cite{Holgado2018}, \cite{Holgado2019} and \cite{SimonDiaz2024}. The remaining stars are considered likely single (LS), although it is possible that some may still harbor undetected binary companions.

For all 358 stars in the working sample, robust estimates of \Teff\ and \grav\ are available (see Sect.~\ref{section31}).
However, as described Appendix~\ref{AppCuts} (see also Sect.~\ref{section32}), we decided to exclude 124 of these stars from what we consider our final "calibrator sample" since they were not fulfilling several quality checks we applied to identify stars with unreliable distance estimates or photometric measurements. We wanted to avoid the potential impact this might have in the reliability of the subsequent determination of the remaining fundamental parameters (i.e., radius, luminosity, and mass).

\begin{figure}[!t]
\includegraphics[width=0.48\textwidth]{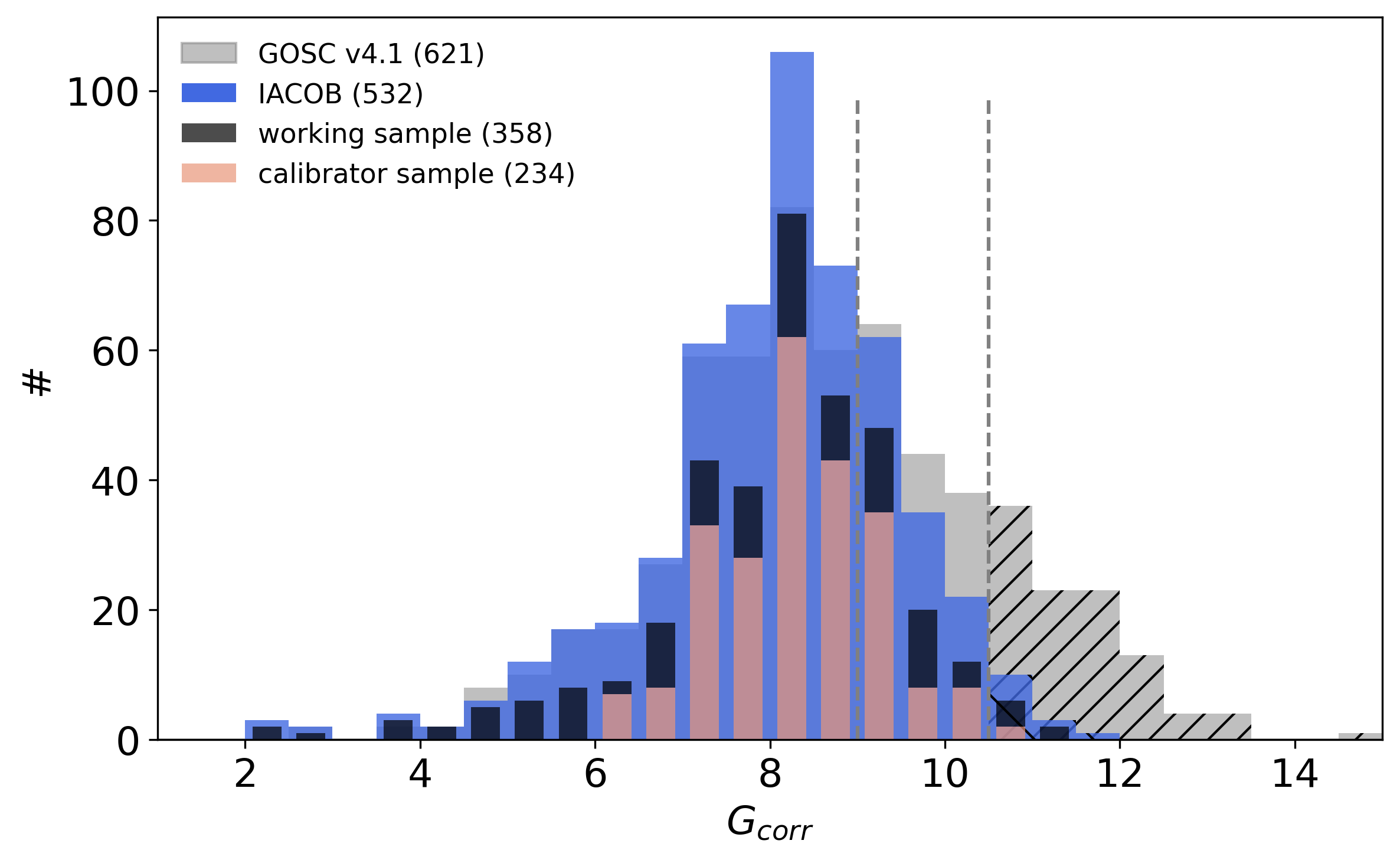}
\caption{
Distribution of the various samples described in Sect.~\ref{section2} as a function of $G_{\rm corr}$, the {\em Gaia G} magnitude \citep[corrected following][]{MaizApellaniz2018b}. Vertical dashed lines indicate the 99\% and 85\% completeness limits of the full IACOB sample, using GOSC v4.1 as a reference. Stars dimmer than $G_{\rm corr}$\,=\,10.5~mag are highlighted with a dash pattern, providing a reference for Figs.~\ref{Hist_SpT_Gaia} and \ref{Hist_d_Gaia}.
}
\label{Hist_G_Gaia}
\end{figure}

\begin{figure}[!t]
\includegraphics[width=0.48\textwidth]{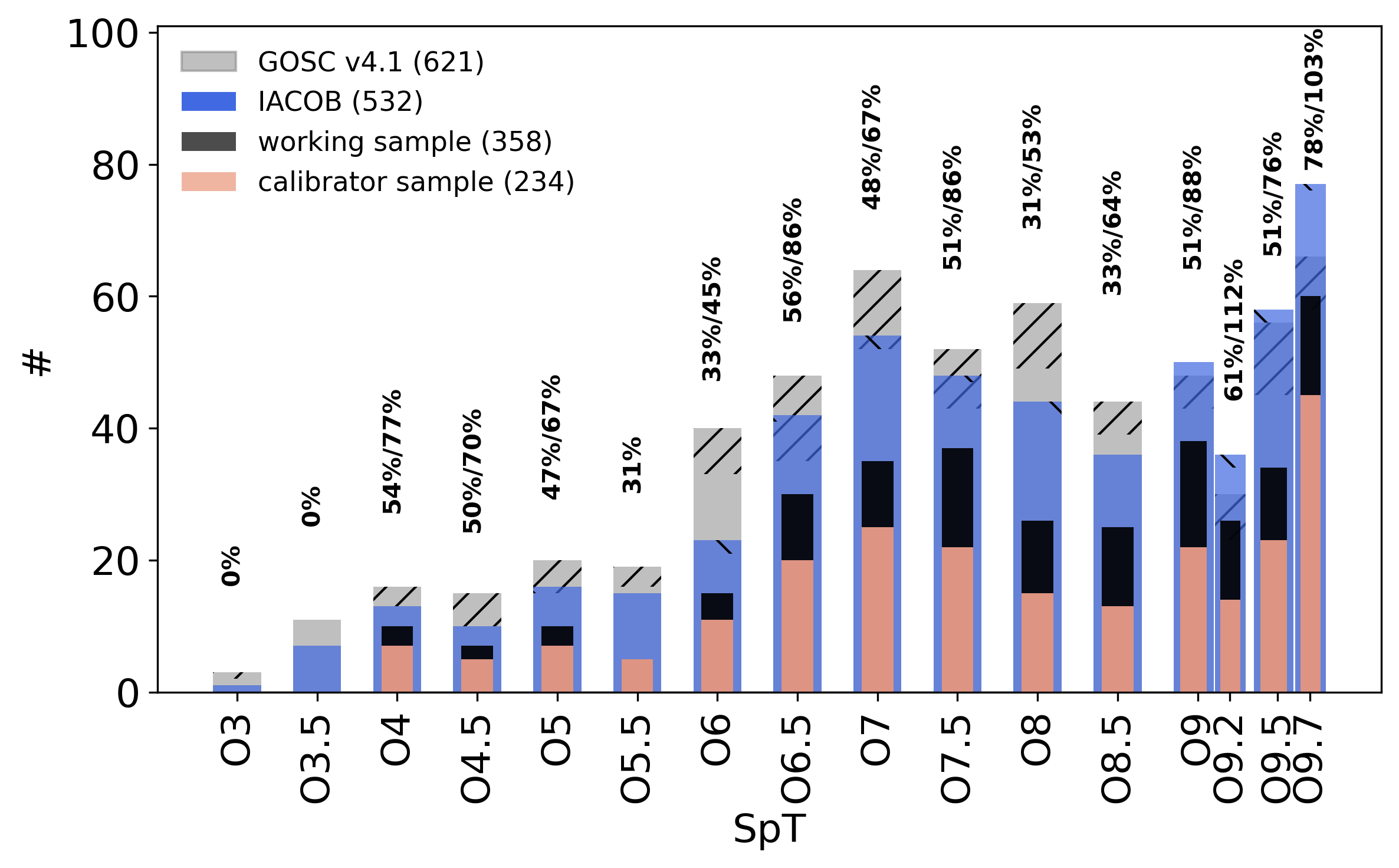} 
\caption{Similar to Fig.~\ref{Hist_G_Gaia}, but showing the distribution of SpTs. Percentages indicate the completeness of stars marked in pink and pink+black with respect to GOSC v.4.1, excluding stars dimmer than $G_{\rm corr}$\,=\,10.5~mag (marked with dash pattern). 
}
\label{Hist_SpT_Gaia}
\end{figure}

\begin{figure}[!t]
\includegraphics[width=0.48\textwidth]{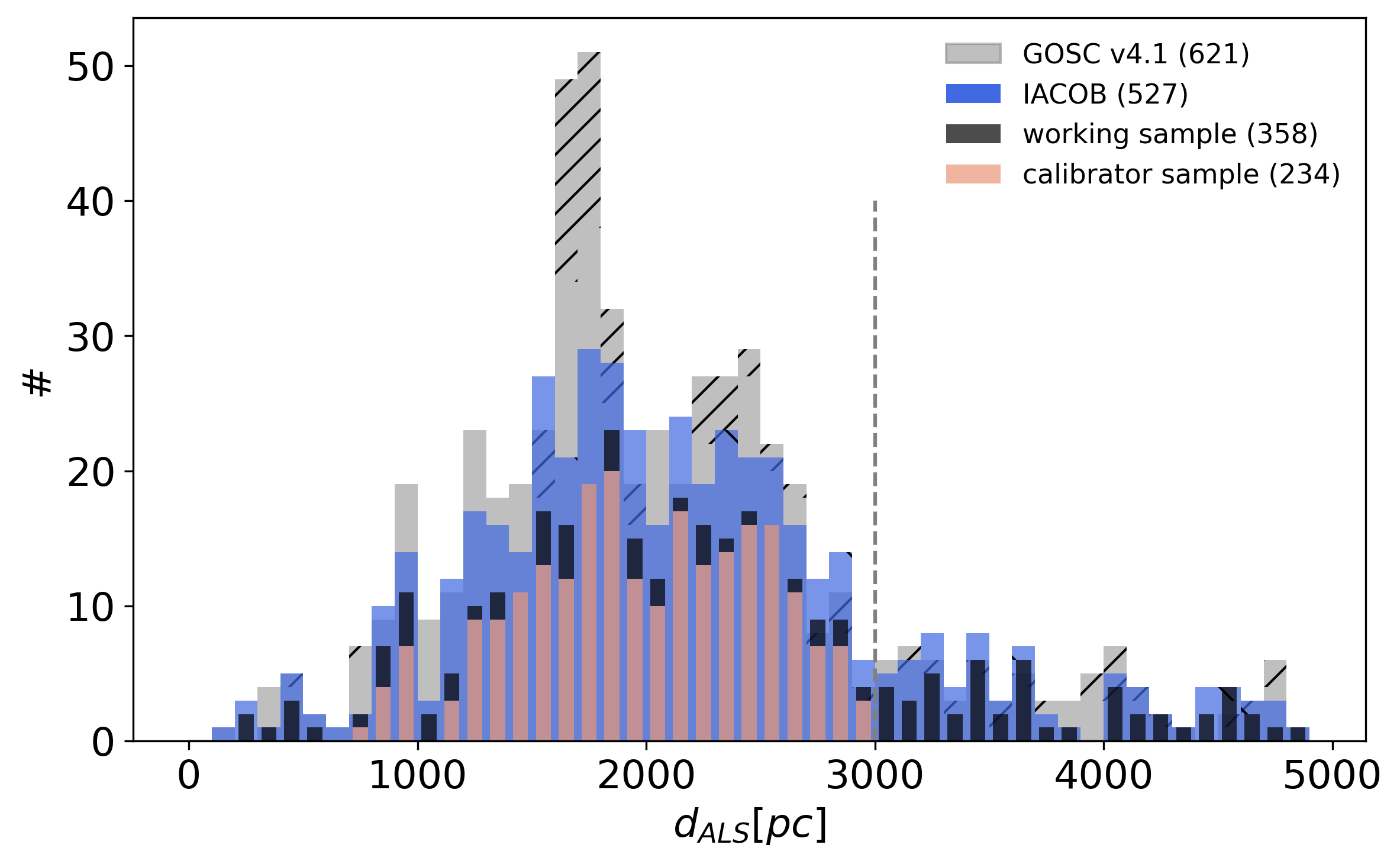}
\caption{Similar to Fig~.\ref{Hist_G_Gaia}, but showing the distribution of distances (from ALS, see Sect.~\ref{section32}). A vertical dashed line indicates the 3000~pc limit used as a quality cut (see Appendix~\ref{AppCuts} for explanation). The large concentration of stars missing in the IACOB spectroscopic database at $\sim$1600 pc corresponds to relatively highly extinguished O-type stars in the Cygnus\,OB2 region \citep[see][]{Berlanas2020}. 
}
\label{Hist_d_Gaia}
\end{figure}

Figures~\ref{Hist_G_Gaia}, \ref{Hist_SpT_Gaia}, and \ref{Hist_d_Gaia} summarize the distribution of the sample under study in terms of {\em Gaia\,G} magnitude, SpT, and distance, respectively. The distributions are shown in comparison to the full set of stars quoted in GOSC v4.1 (621, gray) and those with available IACOB spectroscopy (IACOB sample, 532, blue). 
 Within the working sample of \numO\ stars, we highlight in pink the \numCalib\ stars comprising what we define above as {calibrator sample}, while the remaining \numElim\ are seen in black.

As is further detailed in Appendix~\ref{AppCuts}, the later group primarily includes stars with unavailable or inaccurate {\em Gaia} distances, along with a few cases of poor-quality photometry. For these stars, the absolute V magnitude  estimates are unreliable, which propagates unconstrained uncertainties into the derived luminosities, radii, and masses. Nevertheless, since we successfully performed a quantitative spectroscopic analysis for all of them, we retain this sample to investigate the intrinsic scatter in the SpT\,--\,\Teff\ and \grav\ relations. 
In a forthcoming paper (Holgado et al., in prep.), we plan to leverage two \Mv\ calibrations derived from this {calibrator sample} 
(see Sect.~\ref{Application})
to estimate the remaining fundamental parameters for this discarded sample.

An examination of the three figures illustrates how representative the complete IACOB sample is, at present, with respect to the stars in GOSC v4.1. 
We emphasize that the reference version of GOSC employed in our work is regarded as complete down to B = 8. Beyond that limit, it still includes stars as faint as B = 16, although with a progressively lower level of completeness compared to the number of stars expected per magnitude interval \citep[see comments in][]{Holgado2020}.
Particularly, we consider that for stars brighter than $G$\,=\,10.5\,mag, we have achieved a completeness level of 85\%. 
Notably, given the sample size and the absence of any significant observational bias (aside from that imposed by the brightness of the stars), we have successfully populated all SpT bins, even when considering only the more limited {calibrator sample}. Additionally, our working sample of \numO\ stars encompasses nearly the entire known population of non-SB2 or peculiar O-type stars in the Milky Way within a distance of $\sim$3\,kpc.

\section{From observations to fundamental parameters}\label{section3}

\subsection{Quantitative spectroscopic analysis: \Teff, \grav, and \vsini}\label{section31}

Details on the spectroscopic dataset and analysis strategy to obtain estimates of the effective temperatures and surface gravities (\grav) can be found in \citet{Simon-Diaz2011, SimonDiaz2020a} and \citet{Holgado2018,Holgado2020}. In particular, we have recently repeated the quantitative spectroscopic analysis of the 285 stars from \citet{Holgado2018,Holgado2020} using the same \fastwind\ model grid \citep{Santolaya-Rey1997, Puls2005} but with finer sampling in helium abundance -- our only modification -- motivated by the need for more precise $Y_{\rm He}$ estimates \citep{MartinezSebastian2024}. The same methodology, based on the use of the {\sc iacob-gbat} tool \citep{Simon-Diaz2011}, was applied to an additional 115 stars added to the IACOB database since \citet{Holgado2020}, analyzed directly with the extended grid.
Projected rotational velocities (\vsini) were obtained with {\sc iacob-broad} \citep{Simon-Diaz2014}, following \citet{Simon-Diaz2007,Simon-Diaz2017}, and many values were already published in \cite{Holgado2022}.

\subsection{Distances and extinction}\label{section32}

Thanks to the {\em Gaia} mission, we now have access to parallax measurements for significantly larger samples of Galactic O-type stars than ever before. Notably, 515 of the 532 O-type stars currently included in the IACOB spectroscopic database have corresponding entries in the latest version of the catalog by \citet[][hereafter BJ21]{Bailer-Jones2021}, which serves as a key reference for {\em Gaia}-based distance estimations.

As has long been recognized \citep{LutzKelker1973}, estimating distances by simply taking the inverse of the observed parallax leads to biased or even absurd results, particularly when observed parallaxes are small or negative. Consequently, more sophisticated methodologies are required to transform parallaxes into reliable distance estimates \citep[e.g.,][]{MaizApellaniz2005, BailerJones2018, Anders2019, Pantaleoni2021,MaizApellaniz2021}.

In this work, we utilized two distinct sources of distance estimates. First, we cross-matched our working sample of \numO\ O-type stars with BJ21. Specifically, we adopted distance estimates derived solely from astrometric data, as astrophotometric distances can be affected by saturation and other photometric issues, particularly in the case of blue massive stars \citep{Martin-Fleitas2014, MaizApellaniz2017}. Second, we utilized private data \cite[provided within the framework of the ALS project; see][]{Pantaleoni2021} shared by our collaborators J. Ma\'iz Apell\'aniz and M. Pantaleoni Gonz\'alez. They have developed a specialized method to convert parallaxes into distances specifically tailored for Galactic O- and B-type stars. As is described in \citet{MaizApellaniz2021}, their approach incorporates subtle zero-point corrections and applies a Galaxy model prior optimized for the spatial distribution of massive stars \citep{MaizApellaniz2008}. 
Given their improved treatment of zero-point corrections and optimized priors for massive stars, ALS III distances \citep{Pantaleoni2025} were adopted for the rest of the analysis, while discrepancies with BJ21 estimates were used as a filtering criterion.

For a detailed explanation of the quality filters applied to identify stars with the most reliable distance measurements, ultimately forming the {calibrator sample}, we refer the reader to Appendix~\ref{AppCuts}. Among these criteria, one considered filter is the consistency between the two sets of distance estimates described above, which we quote as $d_{\rm BJ}$ and $d_{\rm ALS}$, respectively. Additional filters include the exclusion of extremely bright stars, for which the standard reduction pipeline of {\em Gaia} data does not yield reliable results, stars located farther than 3~kpc, and those with relative uncertainties $>$30\%. After applying these filters, a total of \numCalib\ stars were identified as having robust distance estimates. 

We also estimated the extinction in the V band ($A_{\rm V}$) for each star in our {calibrator sample}. To achieve this, we compiled a photometric dataset spanning multiple bands, from the optical to the near-infrared, ensuring uniformity by selecting photometric sources with the highest internal consistency and minimal saturation issues. Specifically, $B$ and $V$ magnitudes were obtained from \citet{Mermilliod2006}, while $J$, $H$, and $K_{\rm s}$ magnitudes were sourced from the online 2MASS catalog \citep{Skrutskie2006} via Vizier. Extinction estimates and their associated uncertainties were then calculated using a suite of IDL programs developed by M. A. Urbaneja, following the methodology described in \citet{Urbaneja2017}. 
Their approach is based on fitting the observed spectral energy distribution to a grid of theoretical templates, iteratively determining the combination of extinction parameters that best reproduces the compiled multiband photometry.
For this study,  we adopted the extinction laws proposed by \citet{MaizApellaniz2014}, which provide improvements over the widely used \citet{Cardelli1989} extinction laws.

As is detailed in Appendix~\ref{AppCuts}, we conducted additional sanity checks to identify and exclude unreliable photometric measurements that could result in unconstrained $A_{\rm V}$ estimates. This process further refined our {calibrator sample} by removing stars with problematic photometry. Remarkably, only 8 stars in our working sample of \numO\ failed to meet these quality criteria.

\subsection{Absolute magnitudes, radii, luminosities, and masses}\label{section33}

For all stars with available photometry and reliable distance and extinction estimates, we calculated the absolute V magnitude using the standard equation:
\begin{equation}\label{eq1}
M_{\rm V} = 5 - 5 \log d + m_{\rm V} - A_{\rm V}.  
\end{equation}

\noindent
Subsequently, radii ($R$), luminosities ($L$), and spectroscopic masses ($M_{\rm sp}$) were computed for the entire sample using the following equations:

\begin{equation}\label{eq2}
5 \log(R/R_{\odot}) = 29.57 - M_{\rm V} + V_{\rm syn},
\end{equation}

\vspace{-0.55cm}

\begin{equation}\label{eq3}
\log(L/L_{\odot}) = 2\,\log (R/R_{\odot}) + 4 \log T_{\rm eff}  - 4 \log T_{\rm eff, \odot},
\end{equation}

\vspace{-0.55cm}

\begin{equation}\label{eq4}
\log(M_{\rm sp}/M_{\rm sp, \odot}) = 2\,\log (R/R_{\odot}) + \log g_{\rm true} - \log g_{\odot},
\end{equation}

\noindent
where, Eq.\ref{eq2} was originally introduced by \cite{Kudritzki1980}. Here, $V_{\rm syn}$ represents the wavelength-integrated synthetic emergent flux associated with the best-fitting \fastwind\ model scaled to the V band \citep[see][]{Holgado2019}. Eq.~\ref{eq4} incorporates the surface gravity corrected for centrifugal forces (\gravt), which is derived from the surface gravity obtained from the quantitative spectroscopic analysis using the equation outlined by \cite{Herrero1992} and \cite{Repolust2004}:

\begin{equation}\label{eq5}
g_{\rm true} = g + (v \sin i)^2 / R.
\end{equation}

\noindent
To account for uncertainties in these derived quantities, we performed uncertainty propagation using 5000 Monte Carlo simulations. During this process, we fixed the uncertainties associated with \Teff, \grav, \vsini, $d$, and $A_{\rm V}$ to those obtained from the respective quantitative analyses of spectroscopic, distance, and extinction data. Final values and associated uncertainties for each computed parameter were determined as the median and median absolute deviation of the resulting distributions, respectively.

\section{Results and discussion}\label{section4}

\subsection{Calibrations of fundamental parameters}\label{section41}

\begin{figure*}[!t]
\includegraphics[width=\textwidth]{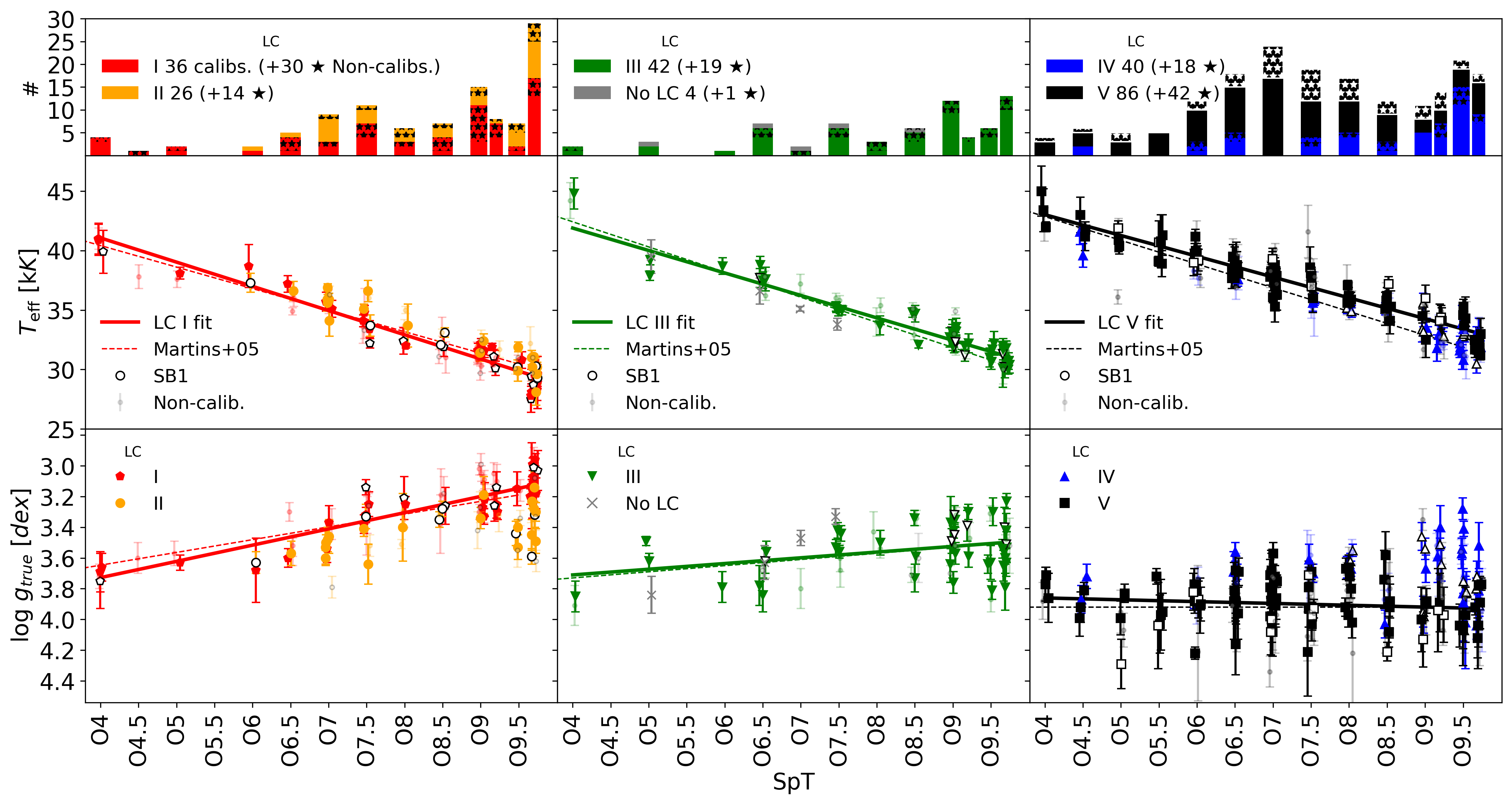}
\caption{SpT\,--\,\Teff\ (middle panels) and SpT\,--\,\gravt\ (bottom panels) calibrations and the \numO\ O-type stars within the working sample, i.e., with available estimations of \Teff\ and \grav. \numCalib\ stars comprising the {calibrator sample} (solid points) and \numElim\ eliminated (semitransparent) due to quality-cut criteria (Appendix~\ref{AppCuts}). Columns divided by LC groups. Dashed lines represent the ``observational scales'' proposed by M05, while solid lines correspond to the calibrations derived in this work (see definitions in Table~\ref{Cab1}). 
Stars identified as SB1 are highlighted with open symbols. 
The top panels display the number of stars per SpT bin, where number in brackets (and hatched bars) indicates stars excluded from the {calibrator sample} (see Sect.~\ref{section2}).
}
\label{Calibs_Tg_calibs}
\end{figure*}

In this section, we focus on exploring the main characteristics of the existing relations between the stellar properties derived for the  \numCalib\ stars in the {calibrator sample} and their SpTs and LCs. It is important to emphasize that all stellar quantities and their associated uncertainties for the stars in this sample have been derived directly, without relying on any pre-existing empirical calibration. Furthermore, this sample has passed rigorous quality control filters, ensuring the exclusion of stars with unreliable data related to spectroscopic analysis, distance, or photometry, thus providing a high-confidence dataset.

Notably, this sample represents a significant improvement over those used in previous studies, both in terms of size and the characterization of spectroscopic binarity. For example, the widely cited work by M05, which serves as a consolidated reference, included only 45 stars and did not account for the potential SB1 status of some stars. 
While our work does not introduce any substantial modeling improvement, \citep[as was, e.g., the consideration of line-blanketing effects in][]{Martins2005},
in contrast, our dataset benefits from a homogeneous and carefully revised spectral classification by the GOSSS team \citep[e.g.,][]{Sota2011, Sota2014, MaizApellaniz2016}, along with uniformly high-quality \textit{Gaia} distances across the full sample. These two aspects further enhance the robustness and consistency of the derived calibrations, particularly for distance-dependent fundamental parameters.

\begin{figure*}[!t]
\includegraphics[width=\textwidth]{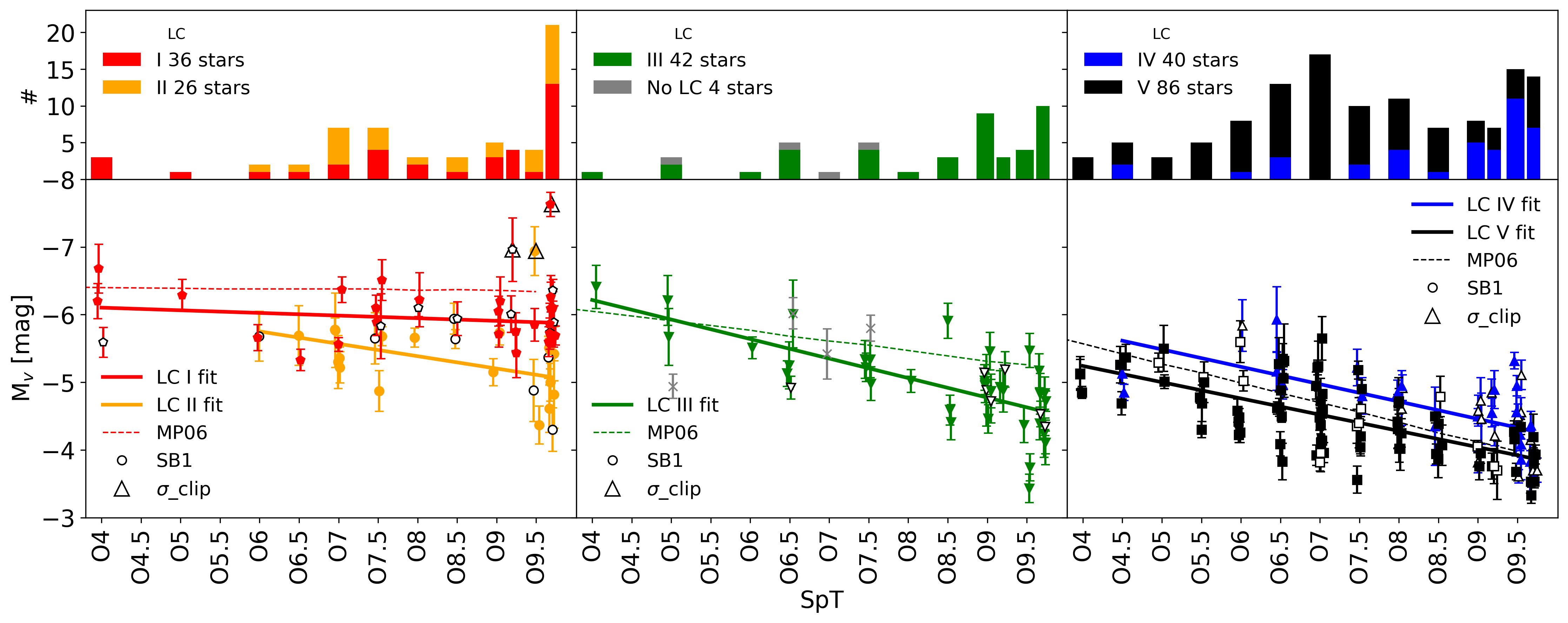}
\caption{SpT\,--\,\Mv\ calibrations for the \numCalib\ stars in the {calibrator sample}, with columns divided by LC groups. 
The top panels display the number of stars per SpT type bin. 
Dashed lines represent the synthetic photometry calibrations proposed by \citet{Martins2006}, while solid lines correspond to the calibrations derived in this work (see definitions in Table~\ref{Cab2}).
A triangle over the symbol indicates stars excluded from the fit by the $\sigma$-clipping procedure.
}
\label{Calibs_Mv}
\end{figure*}

Table~\ref{tableValues} (full version available at the CDS) presents the complete set of empirical data compiled for the {calibrator sample}, serving as the foundation of this study. While some of this information has been previously published \citep[e.g.,][]{Holgado2018, Holgado2020} or compiled from literature, it is included here for the sake of completeness and to consolidate all relevant data into a single source. This table is further supported by several figures (Figs.~\ref{Calibs_Tg_calibs} to \ref{Calibs_RL_calibs}) and additional tables (\ref{tableValues_LC_Teff} to \ref{tableValues_LC_Msp}), which provide detailed insights into the investigated relations.

For the generation of the final calibrations, the {calibrator sample} is divided into three distinct subsamples based on LC: one for dwarfs (LC V), another for giants (LC III), and a third encompassing supergiants (LC I). A calibration with respect to SpT was derived for each of these LC subsamples using linear regression, incorporating sigma-clipping and bootstrapping, while excluding uncertainties from the fits. Although results for stars classified as subgiants (LC IV) and bright giants (LC II) are also displayed in the figures, they were not used to define the aforementioned calibrations. The figures additionally show the number of stars per SpT bin, highlight stars identified as SB1, and provide comparative insights by incorporating the calibrations proposed by M05.

We now examine each of the revised calibrations individually. We highlight the main trends and note significant differences with respect to the reference calibrations by M05.

\subsubsection{Effective temperature and gravity}

For the \Teff\ and \grav\ calibrations, the sanity checks performed to ensure the reliability of the distances and photometry of the {calibrator sample} are not critically important.
However, the photometric data remain indirectly relevant, as they can indicate the presence of companion stars that may affect the reliability of the calibrations.
Notably, the calibrations presented in this work are based on the analysis of optical spectra, and we acknowledge that \Teff\ may differ when derived from other wavelength ranges \citep{Garcia2004,Heap2006,Bouret2013}. This discrepancy likely arises from the sensitivity of UV diagnostics to parameters beyond \Teff, such as wind clumping or the inclusion of X-rays. These factors may affect ionization in the wind but not in the photosphere, thus altering UV line strength without influencing the optical spectrum. Our approach preserves observational consistency and is consistent with previous studies that underscore the reliability of optical diagnostics.

Unlike previous studies based on the use of theoretical models, the consideration of observational samples introduces some limitations regarding the number of available targets in some SpT bins, particularly evident in our case for giant stars. However, our sample size remains robust enough for meaningful statistical analysis. The influence of metallicity on our results is negligible due to the homogeneous composition of the sample. 

Figure~\ref{Calibs_Tg_calibs} illustrates the relationship between \Teff\ and SpT for different LCs.
For the dwarf sample, the linear regression applied to reliable stars shows slightly more scatter than for LC I and III, although the sigma-clipping method used in the regression does not identify outliers.

Figure~\ref{Calibs_Tg_calibs} also extends the analysis to the surface gravity (corrected by the effect of centrifugal forces, see Eq.~\ref{eq5}).
A strong correlation between log \(g\) and LC is particularly pronounced for late SpTs, a well-known trend that is clearly reaffirmed in this study. 
In contrast, the SpT–\grav\ calibrations for all LCs tend to converge toward a similar value for the early O-type stars ($\sim$3.7 dex).
Although the linear approximation for log \(g\) proves generally reliable, it exhibits significantly greater scatter than that observed for \(T_{\text{eff}}\), especially in LC V \citep[see further discussion about this result in][]{SimonDiaz2014a}. 
A slight limitation arises from the underrepresentation of LC I and III stars at SpTs earlier than O6, but the few available examples deviate little from the expected linear trend.

All panels in Fig.~\ref{Calibs_Tg_calibs} also includes semitransparent markers for the \numElim\ stars excluded during quality checks (indicated in parentheses in the histograms), and white symbols for the 82 SB1 stars. 
No noticeable differences are observed in the distribution of either excluded stars or SB1 systems relative to the rest of the sample, in terms of \Teff\ or \grav.

Lastly, to comment on the stars that deviate most from expectations: two non-calibrator stars classified as LC V (ALS~8294 and BD+~453216~A) appear as outliers in \Teff. A visual inspection of their spectra revealed no clear anomalies, suggesting that their spectral classifications may require revision. 
Additionally, one calibrator star (BD$-$~164826, O5\,V, SB1) shows an unusually high gravity ($\log g \sim 4.3$~dex). Although no SB2 signature is evident in our spectra, \citet{MaizApellaniz2019} reported subtle SB2 features, suggesting that contamination from the companion in the Balmer lines may be biasing the gravity estimate.

\subsubsection{Absolute V magnitude}\label{Mvcalib1}

\begin{figure*}
\includegraphics[width=0.98\textwidth]{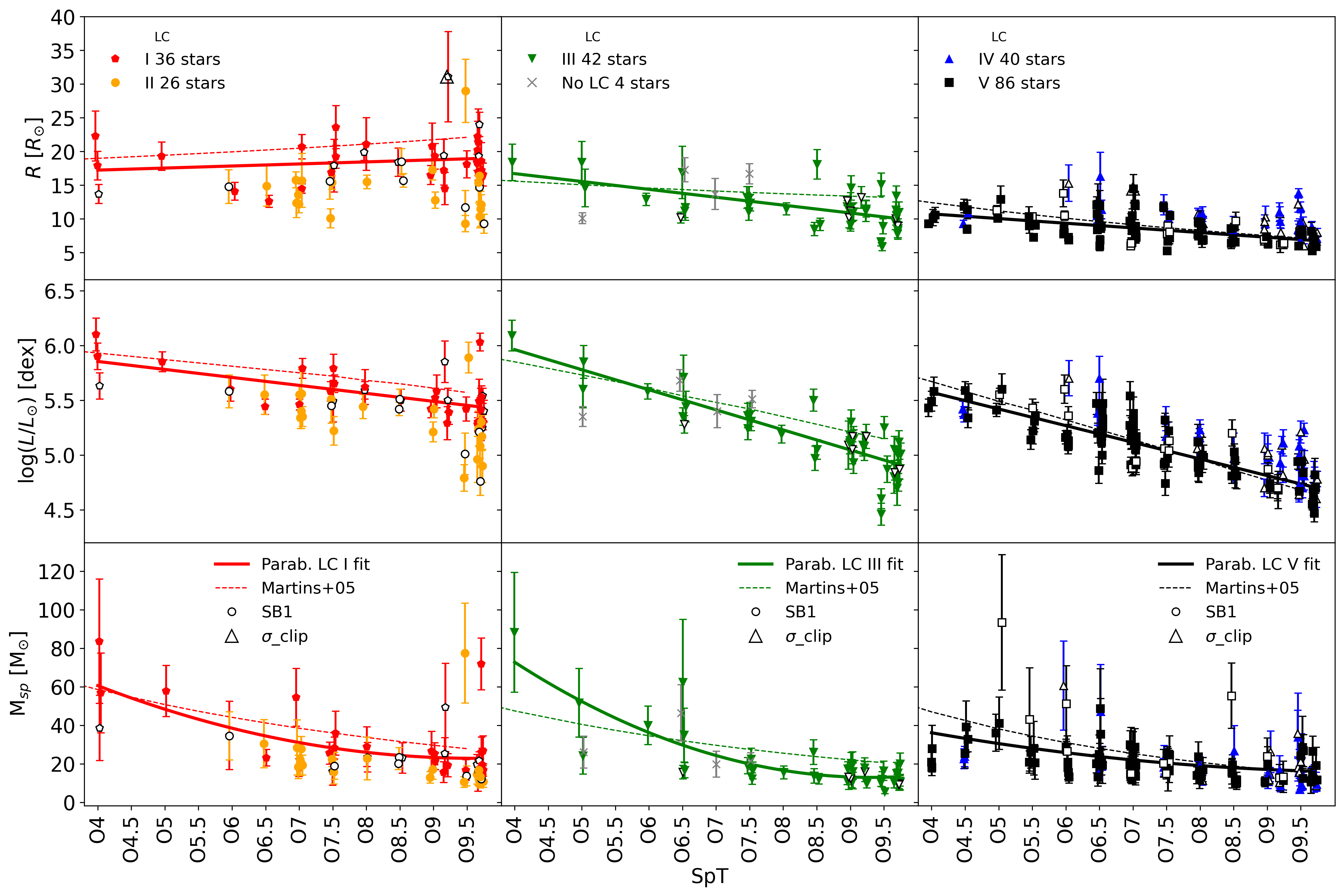}
\caption{SpT\,--\,R (upper panels), SpT\,--\,\Llum\ (middle panels),  and SpT\,--\,\Msp\ (bottom panels) calibrations for the \numCalib\ stars in the {calibrator sample}, with columns divided by LC groups. 
The values were derived from observed absolute magnitudes obtained with distance values and limits from the ALS III work; see Section~\ref{section2}.
The error bars take into consideration the propagation of errors in distances, magnitudes, extinction, and spectroscopic parameters.
Dashed lines represent the observational scales proposed by M05, while solid lines correspond to the calibrations derived in this work (see definitions in Tables~\ref{Cab3} and~\ref{Cab4}).
A triangle over the symbol indicates stars excluded from the fit by the $\sigma$-clipping procedure.}
\label{Calibs_RL_calibs}
\end{figure*}

As is illustrated in Fig.~\ref{Calibs_Mv}, a loose correlation between \Mv\ and LC can be seen for later SpTs, although the scatter is significant.  For early types, the dependence is even weaker, though still suggested. Notably, for stars classified as supergiants, \Mv\ remains nearly constant across SpTs. The linear fit for \(M_V\) is generally robust and consistent under bootstrap analysis, despite notable scatter across all three LCs. 
Additionally, we found that for \(M_V\) most of the \numElim\ stars excluded based on quality criteria deviate significantly from the calibrations, highlighting the robustness of the selection process.

Historically, the large observed scatter in $M_V$ for O-type stars was often attributed to significant uncertainties in distance estimates \citep[e.g.,][]{Moffat1998,Schroder2004,Martins2005}. In our study, we effectively remove distance uncertainty as a major source of dispersion. 
However, despite the use of high-precision \textit{Gaia} distances, the scatter in $M_V$ remains large, suggesting that intrinsic and systematic effects dominate the observed dispersion.
As is shown in Table~\ref{tableValues_LC_Mv}, most SpT bins exhibit dispersions above 0.2~mag with an average near 0.23~mag, significantly lower than the values reported by M05. Notably, increasing the number of stars per bin does not reduce the scatter -- in fact, in some cases it increases -- suggesting that the dispersion is not purely statistical. 
This points toward a possible intrinsic spread in $M_V$, due to a stronger-than-anticipated dependence on secondary effects such as undetected or unresolved binarity, misclassification in LC, or even orientation effects. 
In particular, inclination-driven gravity darkening due to moderate or high stellar rotation could alter the observed flux distribution and inferred magnitudes \citep[see e.g.,][]{vonZeipel1924, Lamers1997, Abdul-Masih2023}.

Stars identified as SB1 are scattered without any clear pattern relative to the calibration. A group of LC~IV–V stars with SpT O6, stand out a little brighter than the calibration value. These are discussed further in Sect.~\ref{LasSB1}.

Three LC I and II stars in the O9\,–\,O9.7 range show significantly brighter $M_V$ values than the rest of the sample, being clear outliers from the trend. These include HD\,195592 (O9.7\,Ia$+$, \Mv\,=\,-7.63) and HD\,152424 (OC9.2\,Ia, \Mv\,=\,-6.96), both likely affected by their extreme luminosity Ia subclass, as noted by \citet{Evans2004}. 
Another outlier is HD\,190429\,B (O9.5\,II–III, \Mv\,=\,-6.94), and its designation suggests a possible unresolved companion, which may be affecting its brightness. Although part of the {calibrator sample}, they were all excluded from the fit via $\sigma$-clipping.
In the LC III group, two ON9.5\,IIIn stars (HD~117490 and HD~91651) appear significantly fainter than the rest. 
The fainter appearance of these stars may be due to strong gravity darkening, which occurs in rapidly rotating stars, specially when viewed equator-on (consistent with their high \vsini).

The values of these discrepant cases resemble stars excluded during the quality filtering process, suggesting that the remaining outliers may also be affected by undetected issues such as binarity, contamination, or misclassification. Continued refinement of SpTs and improved detection of binary companions will be essential to reduce residual scatter in future calibrations.

\subsubsection{Radius and luminosity}\label{sectionRL}

As is illustrated in Fig.~\ref{Calibs_RL_calibs}, the linear fit is, overall, a good agreement for the distribution of radius and luminosity, despite moderate scatter. Notably, only two stars were excluded by $\sigma$-clipping when performing the linear fit to the data. 
We note that the observed scatter in $R$ and $\log L$ slightly exceeds the typical uncertainties associated with the estimation of these parameters, which are approximately 10\% and 0.11\,dex, respectively. This is most likely reflecting the underlying intrinsic scatter also found for $M_V$ and $T_{\rm eff}$ in most SpT-bins.

Stars identified as SB1 are again scattered without any clear pattern relative to the calibration. The group of LC\,IV–V stars with SpT O6 previously identified as brighter in $M_V$ also stand out here with clearly larger radii and luminosities when compared to the rest of the sample, as well as the Ia outliers commented before.

The radius and luminosity calibrations are consistent and broadly applicable. Although, individual use should account for the intrinsic scatter.

\subsubsection{Spectroscopic mass}\label{sectionMsp}

In this work, $M_{\rm sp}$ is derived from \gravt\ and $R$, as described in Sect.~\ref{section33}.
The new calibrations, presented in Table~\ref{Cab4}, are based on the {calibrator sample} and are fitted using a second-order polynomial to account for the curvature observed across the spectral sequence (see Fig.~\ref{Calibs_RL_calibs}). 
This particular fits seems to reproduce the global behavior well, despite moderate scatter. The curvature is most pronounced in the LC~III calibration, likely due to the presence of a single LC~III star with large values of $R$ and \Llum.
For LC~I and LC~V, the derived trends are consistent with those from Martins.

Uncertainties in $M_{\rm sp}$ are relatively large, typically around 30\%, though slightly smaller than those reported in previous works in the literature. The scatter is modest but larger than the one detected in the $R$ and \Llum\,-\,SpT calibrations. This scatter, combined with the persisting large uncertainties associated with the $M_{\rm sp}$ estimates, makes this calibration less reliable for individual mass determinations. As is seen in Table~\ref{Cab4}, average masses per bin align reasonably well with the fit, but the large individual uncertainties caution against overinterpretation. Direct application of \Msp\ calibration is only advised when no gravity measurements are available. Otherwise, combining a calibrated radius or $M_V$ with an observed $\log g$ provides a more precise estimate, as done for the validation exercise described in Sect.~\ref{Application}.

Most SB1 stars do not show significant deviations from the general trend. However, the group of LC~IV–V stars with SpT~O6 previously flagged for their brighter-than-expected $M_V$ values appear here with extremely high $M_{\rm sp}$ estimates (see Sect.~\ref{LasSB1}).

In addition to the outliers already discussed in previous subsections, the most extreme spectroscopic mass values belong to SB1 systems. BD$-$16~4826 (O5V, SB1) exhibits the highest \gravt\ in the sample and correspondingly yields an exceptionally large $M_{\rm sp}$. HD~46149 (O8.5V) shows marginal SB2 indications but is also flagged as a spectral standard, reinforcing the need for accurate characterization of such benchmark objects.

In summary, the spectroscopic mass calibration is overall consistent. However, it is unreliable for individual stars due to large uncertainties and sensitivity to systematics.

\subsection{Comparison with \cite{Martins2005}}\label{MartinsComp}

As is illustrated in Fig.~\ref{Calibs_Tg_calibs}, our empirical \(T_{\text{eff}}\) values align closely with the observational calibrations of M05 for LC\,I and III. However, for LC\,V, our calibration predicts a characteristic \Teff\ approximately 1500\,K higher \citep[see further notes about this discrepancy in][]{SimonDiaz2014a}. 
Regarding \(\log g\), our results agree well with their work in LC III, despite a somewhat large scatter, while small differences (0.1 dex) are observed for LC I and LC V at early SpTs. 
These differences are likely due to the small sample size in these regions, where regression results may be strongly influenced by the limited number of available stars. However, the bootstrap analysis is designed to mitigate such issues effectively.
We must note that we find that our SpT–\grav\ calibration for dwarfs results in gravities identical to the value proposed by M05 (\grav\ = 3.9 dex). However, we remark that assuming a unique value per SpT of this parameter in the O dwarfs is an oversimplified recipe \citep[see e.g.,][]{SimonDiaz2014a,Holgado2018}. 

Compared to M05, our \Mv\ calibrations show a difference of approximately half a magnitude for LC\,I. This is consistent with the standard deviation reported by \citet{Martins2005} for their observational \(M_V\) scale. For LC\,III, the discrepancy is more significant, with our calibrations yielding \(M_V\) values about one magnitude fainter for late SpTs. In contrast, for LC\,V, the differences are smaller: our calibrations align closely with M05 for late SpTs and differ by a maximum of only a quarter of a magnitude for early types. 
These differences can be attributed to the inherent challenges in addressing distance uncertainties and extinction effects in M05 observational sample, reflecting the limitations of the data available at the time.

A comparison with the M05 calibrations for radii and luminosities reveals close agreement for LC V. For LC III, the new calibrations yield smaller radii and luminosities for late SpTs. For LC I, the results consistently show systematically lower radii and luminosities across all SpTs, with average differences of approximately \(4 \, R_\odot\) in radius and \(0.15 \, \text{dex}\) in luminosity.
In M05, luminosities were calculated differently, using \(M_V\) along with the bolometric correction (see further notes in Sect.~\ref{BC}). Despite this methodological difference, the comparison between our results and theirs is remarkably good. 

Regarding spectroscopic masses, our calibrations yield systematically lower masses for LC~I stars, in line with the trends observed in $R$ and log\,$L$. For LC~III, the differences are more pronounced, with our values significantly below those reported by M05. For LC~V, the agreement is generally good, except for early types, where our masses can differ by up to 10\,$M_\odot$.

In summary, our calibrations build on those of M05 using a larger, cleaner sample with precise \textit{Gaia} distances and binarity information, showing systematic \Mv\ differences for LC I–III that affect derived parameters, while late-type LC V stars agree closely. They offer an updated, homogeneous empirical reference.

\subsection{Impact of SB1 systems on the calibrations}\label{LasSB1}

\begin{figure*}[!t]
\includegraphics[width=\textwidth]{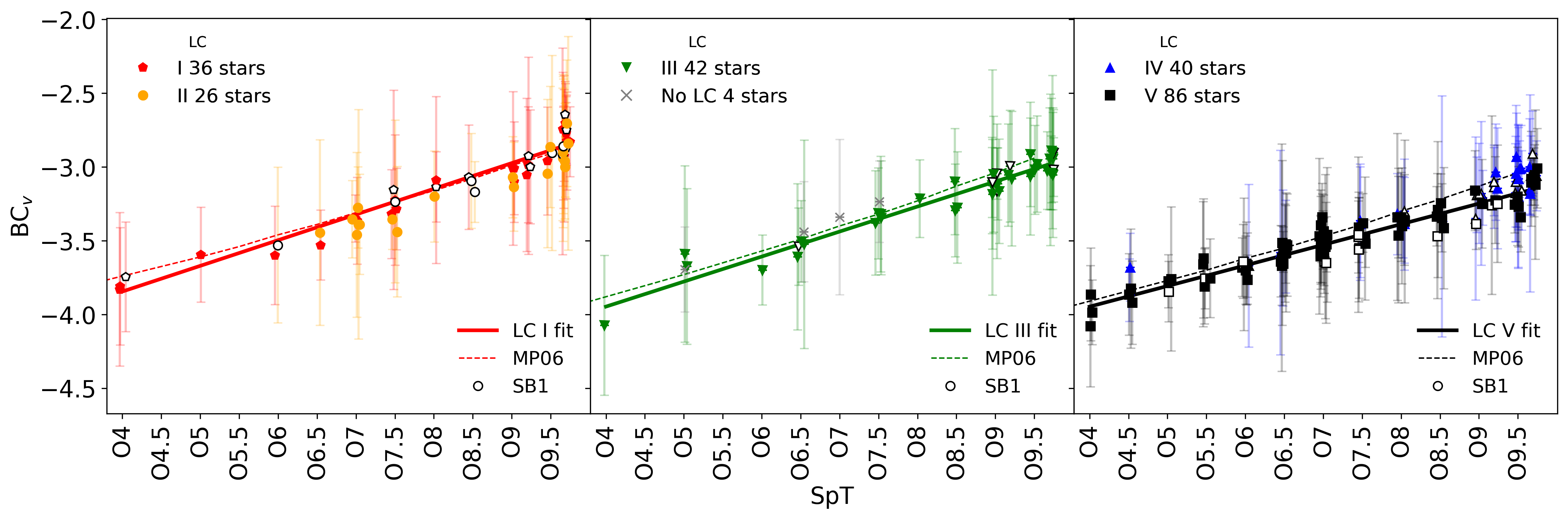}
\caption{SpT\,--\,\BCv\ calibrations for the \numCalib\ stars in the {calibrator sample}, with columns divided by LC groups. 
Dashed lines represent the synthetic photometry calibrations proposed by M05, while solid lines correspond to the calibrations derived in this work (see definitions in Table~\ref{CabBC}).
}
\label{Calibs_BC}
\end{figure*}

Among the \numCalib\ stars in our {calibrator sample}, 52 are classified as SB1. While we have highlighted individual SB1 systems with extreme properties throughout the previous sections, we now assess the global impact of binarity by evaluating how the exclusion of all SB1 stars affects each of the derived calibrations.

In nearly all cases, removing the SB1 subsample does not significantly alter the resulting calibrations. The only notable exception is the spectroscopic mass calibration for LC~V stars (Figure~\ref{Msp_LCV_comp_SB1}), where excluding SB1s results in a slightly steeper trend for early SpTs. Above SpT O6.5, the calibration flattens when SB1s are excluded, with mass differences reaching up to 8~$M_\odot$. This deviation is driven by three SB1 stars with unusually high mass estimates in that regime.

In SB1 systems, an undetected secondary can enhance the total brightness, biasing $M_V$ toward larger radii and \Msp. Additionally, If the companion contributes weakly to the spectral lines, it may artificially broaden the Balmer wings, leading to higher inferred \grav. 
This effect appears to be more pronounced in LC~V stars, likely due to both statistical and physical reasons. First, the LC~V group contains more stars overall, increasing the chance of unresolved systems. Second, the combination of two LC~V stars may result in blended lines that mimic single broadened profiles, making the system harder to detect as SB2. In contrast, in LC~I or LC~III binaries, differences in brightness and line profiles between components make binarity easier to identify, reducing the chance of such systems contaminating the calibrator sample.

We therefore advise caution when applying our calibrations to SB1 stars. If the derived $\log g$ is unusually high, it may be more reliable to estimate $M_{\rm sp}$ using the direct calibration rather than combining $\log g$ with $R$ or \Mv\ calibrated values.

Finally, we recall that \citet{Holgado2022} reported a statistically significant difference in the $v \sin i$ distribution between SB1 and non-SB1 stars in the O-type domain. However, in the context of this study, the inclusion of SB1 systems does not introduce systematic deviations in most calibrations, aside from the mass calibration for LC~V stars, discussed above. 
This supports their inclusion in the computation of the calibrations, as they do not appear to introduce significant contamination.

\subsection{Additional calibrations}\label{section42}

\subsubsection{Bolometric correction}\label{BC}

The bolometric correction (BC) quantifies the difference between the bolometric magnitude of a star and its magnitude in a given photometric band, accounting for the flux emitted outside that passband in the absence of interstellar extinction. In the classical standard mode, one can define a distinct BC for each photometric filter. However, the literature highlights important conceptual and methodological caveats regarding the use and interpretation of bolometric corrections -- particularly the need to avoid inconsistencies across filter systems or the derivation of nonphysical BC values. We refer the reader to the comprehensive discussion in \citet{Eker2025} for a critical overview of these paradigms.

In most works, BC values are derived from synthetic stellar atmosphere models, which are then used to convert absolute magnitudes to bolometric magnitudes, and consequently to luminosities. In contrast, our approach follows the inverse path: having already determined stellar luminosities empirically via the Stefan–Boltzmann law (Sect.~\ref{section42}), we compute the bolometric magnitudes directly, and then infer the visual-band correction \BCv\ as the difference between $M_{\rm bol}$ and $M_V$:
\BCv=M$_{bol}$-\Mv

Figure~\ref{Calibs_BC} displays the empirical \BCv\ values obtained for the {calibrator sample}, separated by LC. Linear fits were performed for each LC group individually. The trends are remarkably consistent across LC I, III, and V, showing nearly identical slopes and intercepts. This supports the long-standing assumption that BC is primarily a function of $T_{\rm eff}$ and, to first order, independent of LC -- an assertion commonly adopted in the literature \citep[e.g.,][]{Martins2005}. 
Despite the minimal differences, separate regressions are provided per LC to preserve internal consistency with other calibrations presented in this work.

Although the errors associated with our BC values are relatively large (typically $\sim$0.35 mag), owing to the propagation of uncertainties from luminosity and distance, the internal scatter around the linear trends is surprisingly low. This suggests that the method, while observationally driven, yields robust and consistent results.

To further assess the validity of our empirical corrections, we compared our \BCv\ values with those predicted by the $T_{\rm eff}$–BC relation provided by M05. Their calibration is based on model atmospheres and synthetic photometry, and assumes a one-to-one relation between BC and $T_{\rm eff}$, independent of LC. As is shown in Fig.~\ref{BCvsMartins}, the agreement is excellent: the differences between the two approaches are well within the uncertainties of our measurements, with no significant systematic offset. This result reinforces the use of a $T_{\rm eff}$-based bolometric correction as a practical approximation, even when derived from models.

\subsubsection{Shape of H$\beta$ as a diagnostic tool in O stars: FW3414}\label{FW3414}

\begin{figure}
\includegraphics[width=0.5\textwidth]{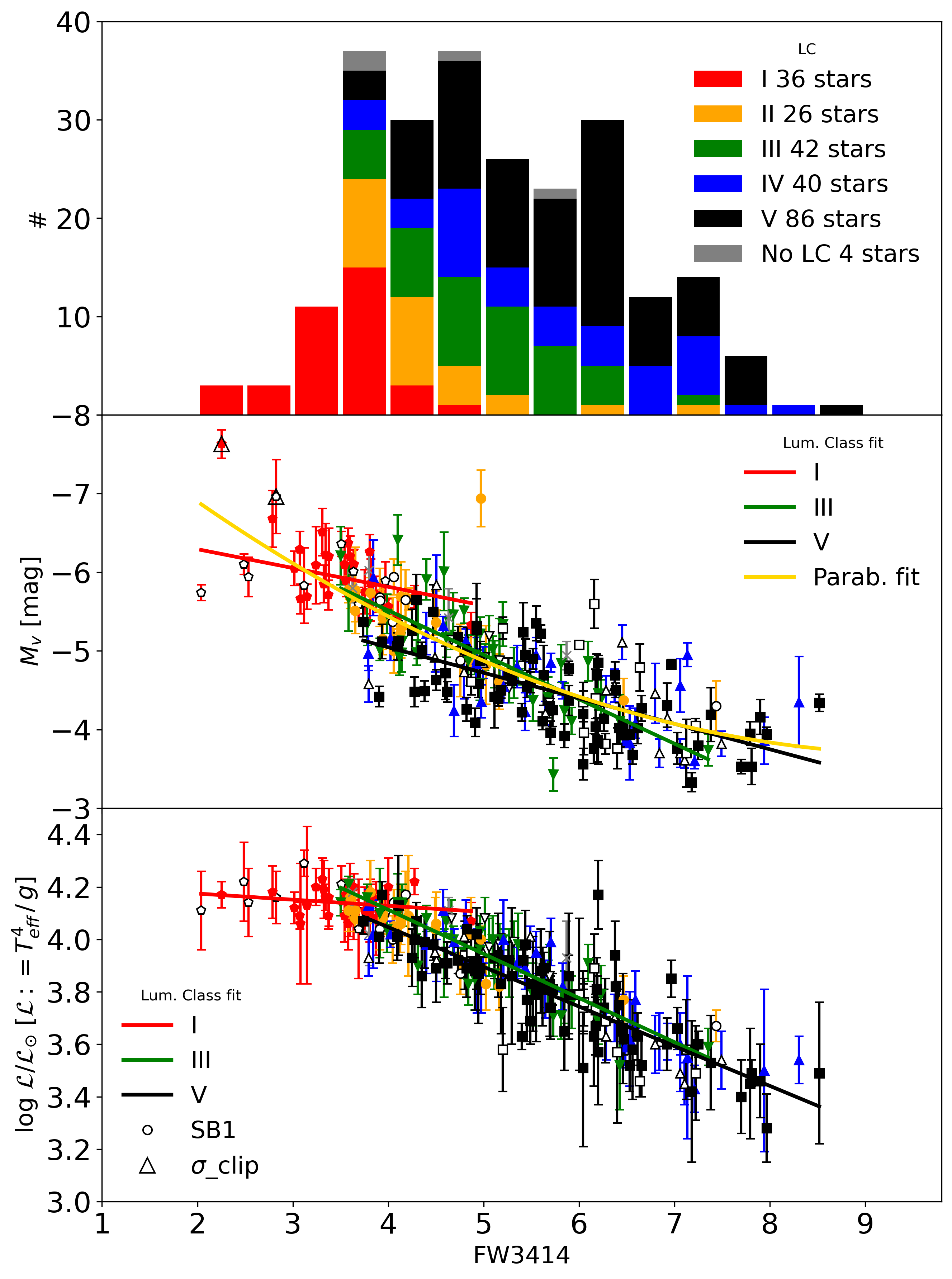}
\caption{FW3414--\Mv\ (middle panel) and FW3414--\Lsp\ (bottom panel) calibrations for the \numCalib\ stars in the {calibrator sample}.
Colors distinguish between the different LCs.
The top panel display the number of stars per FW3414 type bin.
Solid lines indicate the linear fit calibrations derived in this work for each LC.
The solid gold line in FW3414--\Mv\ represents the parabolic fit to the entire sample.
All calibrations are included in Table~\ref{Cab5}.}
\label{Calibs_FW3414_calibs}
\end{figure}

\begin{figure*}[!t]
\includegraphics[width=\textwidth]{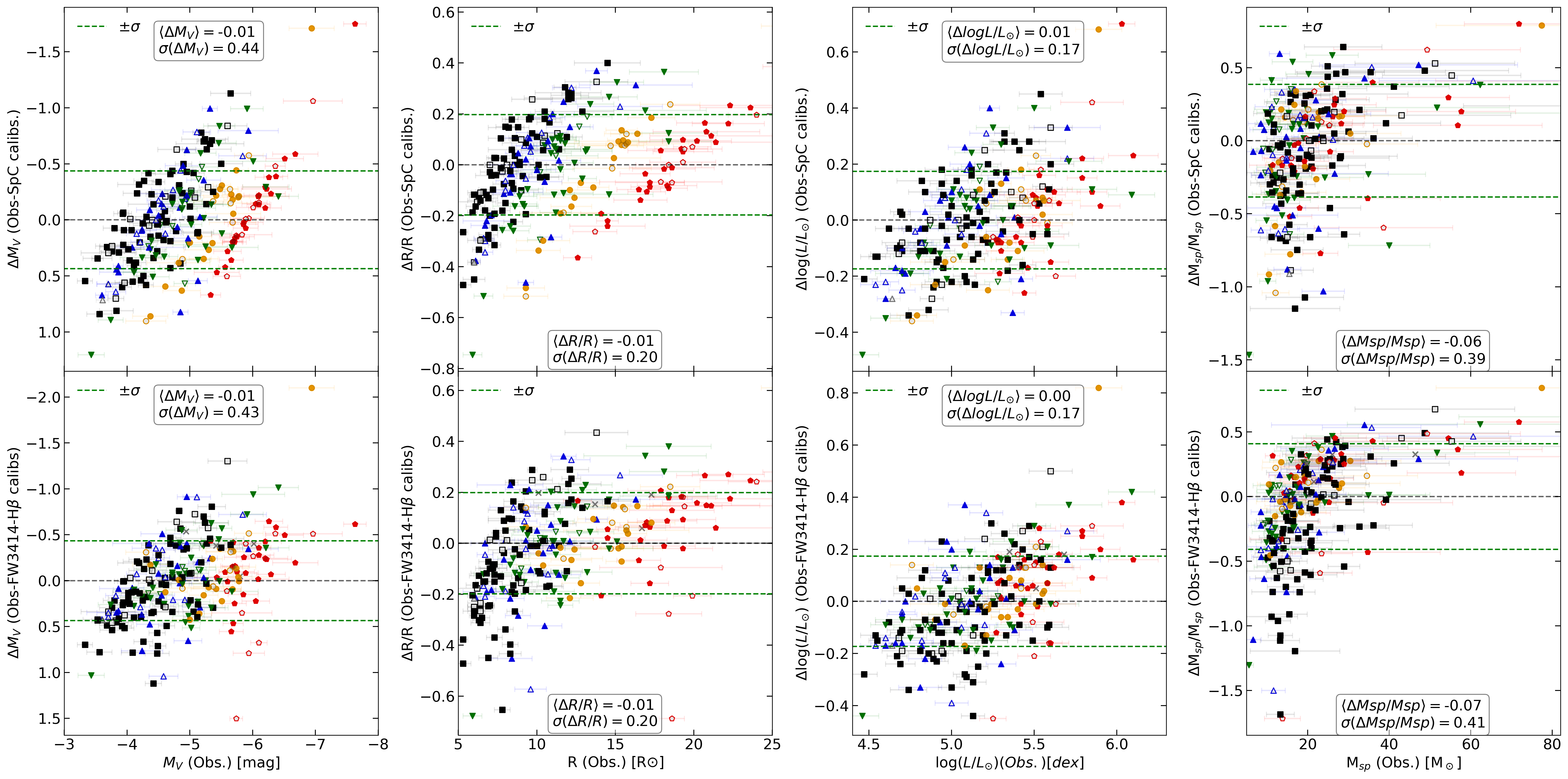}
\caption{
Comparison between observed and calibrated stellar parameters for the \numCalib\ stars in the {calibrator sample}, based on the calibrations of $M_V$ and subsequent derivation of radii, luminosities, and spectroscopic masses (see Sect.~\ref{Application}). Results are shown for both the SpC-based (upper row) and FW3414-based (lower row) calibrations of $M_V$. Each column corresponds, respectively, to the differences in $M_V$, radius, luminosity, and spectroscopic mass. Differences in radius and mass are expressed as fractional values relative to the observed quantities. The LC is color-coded as in previous figures; SB1 systems are marked with white-filled symbols. Dashed green lines indicate the $\pm1\sigma$ region derived from the scatter of the differences. Each panel also includes the mean offset and standard deviation of the residuals, providing an estimate of the typical uncertainty associated with each calibration.}
\label{MvRLM_vsMvRLM_forced_Cuadr_Deltas}
\end{figure*}

The precise spectral classifications used in this work rely on the homogeneous framework provided by the GOSSS and ALS surveys. However, such high-quality classifications may not always be achievable in studies involving distant, heavily obscured stars or extragalactic samples. Anticipating these limitations, we expect similar challenges in the analysis of upcoming large-scale spectroscopic surveys such as WEAVE and 4MOST, in which we are actively involved.

A comparable issue arose during the IACOB project when analyzing B-type giants and supergiants. To address it, we employed what we called the FW3414 parameter offering a fast and reliable alternative. As is described in \cite{deBurgos2023}, FW3414 is defined as the difference between the widths of the H$\beta$ line at three-quarters and one-quarter of its depth. This parameter serves as a proxy for \(\log L\) and provides insights into surface gravity. The use of H$\beta$ minimizes the effects of wind emission that can affect H$\alpha$, while the use of difference between the widths mitigates the impact of \(v \sin i\) on the line shape.

Building on its successful application in B stars, we extended the FW3414 study to the O-star regime and assessed again its potential as a diagnostic of LC, and further expanded its use as a calibrator of fundamental parameters.
The results are shown in Fig.~\ref{Calibs_FW3414_calibs}, which presents the distribution of FW3414 against \Mv\ and \Lsp\, separated by LC. The clear separation of LC I stars (with FW3414$<$4~\AA) from LC III and LC V stars demonstrates the parameter’s utility, confirming trends already observed in B-type stars. LCs III and V show overlap: FW3414 values range from 4–6~\AA\ for giants and 4–8~\AA\ for dwarfs, limiting its use as a standalone classifier between LC III and LC V.

In addition, we derived novel empirical calibrations linking FW3414 to both \Mv\ and \Lsp\ using the {calibrator sample}. 
In cases where LC classification is unavailable -- i.e., when the \ioni{He}{ii}$\lambda$4686 line is not covered, preventing a reliable LC assignment \citep[see e.g., Table 5 in][]{Sota2011} -- we also tested a second-order polynomial fit between FW3414 and \Mv\ using the full sample without LC segregation. The resulting fit, listed in Table\,\ref{Cab5}, shows good agreement with the data and is used in Sect.~\ref{Application} to estimate stellar parameters and assess their consistency with the observed values.

We note that FW3414 measurements may be mildly sensitive to spectral resolution and signal-to-noise ratio. To facilitate future applications, we shall make available the spectra of the calibrator stars used in this study, enabling users to reprocess them at their working resolution and construct tailored calibrations as needed.

Several LC I SB1 stars with FW3414 values below 3.5~\AA\ show the largest deviations from the polynomial fit to the full sample. These cases should be interpreted with caution, as unresolved binarity may affect both the photometric measurements and the shape of the line profiles.

These results confirm that FW3414 is a quite useful empirical tool for estimating absolute magnitudes when full spectral classification is unavailable. When both H$\beta$ and \ioni{He}{ii}$\lambda$4686 are present, their combined use enables a rapid and reliable classification, allowing for accurate estimates of \Mv\ and, subsequently, stellar radii, luminosities, and masses. Even in the more limited case where only H$\beta$ is available, the polynomial calibration of FW3414 alone provides a solid approximation. This method is particularly promising  and serves as the cornerstone of our methodology for analyzing spectra from the WEAVE and 4MOST surveys.

\subsection{Testing the Calibrations: \Mv\ from SpT and FW3414}\label{Application}

In the context of upcoming large spectroscopic surveys, \Teff\ and \grav\ will often be available for many stars, while reliable estimates of \Mv\ may not be obtainable for all of them. In such cases, the \Mv\,--\,calibrations proposed in Sects.~\ref{Mvcalib1} and \ref{FW3414} provide a means to derive stellar radii, luminosities, and masses of individual stars.
To assess the reliability and expected accuracy reached in the determination of these parameters when following this methodology, we present below the results of an experiment we performed using our {calibrator sample}. In brief, we obtained \Mv\ estimates for each individual star using the two abovementioned calibrations:
(i) from the spectral type and luminosity class (SpC), and (ii) from the FW3414 parameter, assuming no prior knowledge of LC. Using these $M_V$ values in combination with the spectroscopically determined $T_{\rm eff}$ and $\log g$, we computed stellar radii, luminosities, and spectroscopic masses using Eqs.~\ref{eq2} to \ref{eq4} and the same Monte Carlo approach as the one described in Sect.~\ref{section33} to propagate uncertainties from all relevant parameters. For $M_V$, the adopted uncertainty corresponds to the standard deviation between calibrated and observed values.

Figure~\ref{MvRLM_vsMvRLM_forced_Cuadr_Deltas} summarizes the outcome of this experiment, where the four quantities (\Mv, $R$, log\,$L$, and \Msp) estimated as described above are compared to the values obtained when considering the observationally determined \Mv.
As expected, $M_V$ and $R$ appear more compartmentalized in the SpC-based approach due to its explicit LC dependence, while the FW3414-based method yields smoother trends. Luminosity and mass comparisons are less affected by LC, as they depend directly on $T_{\rm eff}$ and $\log g$.

Overall, the agreement is satisfactory. For $M_V$, the residual dispersion (0.44~mag) is consistent with the Monte Carlo–derived uncertainties and notably larger than the typical observational error (0.22~mag), validating its use in uncertainty propagation. 
To ensure a consistent and realistic uncertainty budget, we adopt the standard deviation of the differences between calibrated and observed values as the representative uncertainty for each parameter. This choice better reflects the intrinsic scatter in the calibrations and aligns well with the results of the Monte Carlo simulations.

For radius and luminosity, the differences are slightly higher than the expected errors (20\% vs. 10\% in radius; 0.17 vs. 0.11~dex in luminosity). Spectroscopic mass shows the largest deviations, with residuals reaching up to 100\% in extreme cases and a typical scatter of $\sim$40\%, exceeding the $\sim$30\% uncertainty from observed values. These discrepancies reflect the high sensitivity of mass to radius and gravity, and the intrinsic degeneracy of mass across LC groups.

Both methods yield consistent trends, with the FW3414-based calibration providing slightly more linear relations, especially between $M_V$ and radius. Most outliers beyond 1$\sigma$ correspond to previously identified problematic stars, such as LC\,I objects with unusually bright $M_V$ or low FW3414 values. No systematic behavior is observed among SB1 systems. A mild bias is observed in mass residuals: at low masses ($<25\,M_\odot$), some calibrated values tend to be overestimated, while for high-mass stars ($>40\,M_\odot$), the observed masses are generally larger than the calibrated ones -- particularly for the FW3414-based method.

\section{Concluding remarks and future prospects}\label{section5}

In this work, we present a new set of empirical calibrations for physical parameters of Galactic O-type stars, built from a carefully selected and homogeneously analyzed sample of calibrator stars. Using updated atmospheric parameters from the IACOB project and homogeneous {\it Gaia} DR3 distances, we derive calibrations for $T_{\rm eff}$, $\log g$, $M_V$, $R$, log\,$L$, \Msp, and bolometric correction as a function of SpT and LC. 
To ensure reliability, we first excluded known SB2 systems and spectroscopically peculiar stars, yielding a working sample that is both clean and representative of the Galactic O-star population, free from evident selection biases. 
This was subsequently refined through additional quality cuts, resulting in the {calibrator sample} used to derive the final calibrations.
These calibrations constitute an updated reference for the community, building upon the widely used results from \citet[][M05]{Martins2005} and providing stronger statistical support.

Key features of our approach include: (i) a much larger stellar sample, (ii) the use of accurate {\it Gaia}-based distances, and (iii) a carefully cleaned sample. In the latter case, we apply rigorous quality checks on photometry, extinction, distance, and binarity status. 

One of the key findings is that the scatter in $M_V$ remains substantial, even when precise {\it Gaia} DR3 distances are used. This residual dispersion -- likely driven by factors such as unresolved binarity, classification ambiguities, or intrinsic stellar variability -- propagates into the derived radii, luminosities, and spectroscopic masses, ultimately limiting the precision of calibrated values.

Our calibrations show overall agreement with those of M05 in $T_{\rm eff}$ and $\log g$. For LC~V stars, however, we identify a slight offset in $T_{\rm eff}$, our values being systematically hotter for later types. Our derived $M_V$ values tend to be slightly fainter, particularly for LC~I stars, which leads to slightly smaller radii and luminosities in that regime. For other LCs, radii and luminosities are nearly identical to those from M05. Similarly, our bolometric corrections—derived empirically from observed luminosities—are in excellent agreement with those computed from models.

Regarding binarity, we highlight that single-lined spectroscopic binaries (SB1) do not introduce significant systematic shifts in most calibrations. However, a subset of LC~V SB1 stars shows abnormally high spectroscopic masses, likely due to undetected SB2 contamination. These objects distort the upper envelope of the mass distribution, and excluding them leads to a slightly lower calibration in that regime. Special care is therefore advised when interpreting unusually high $\log g$ values in dwarfs.

We also present, for the first time in O stars, empirical calibrations for the FW3414 parameter -- derived from the H$\beta$ profile shape -- as a diagnostic of $M_V$ in cases where full spectral classification is unavailable. Originally introduced for B-type stars \citep{deBurgos2023}, this parameter proves effective in separating LCs and estimating some stellar parameters as \Mv\ and \Lsp\ from a single spectral line. A second-order polynomial fit to the full sample (independent of LC) offers a promising alternative for upcoming large-scale spectroscopic surveys (e.g., WEAVE, 4MOST), particularly in cases of limited spectral coverage or when accurate spectral classification is not feasible.

As a consistency test, we compared radii, luminosities, and spectroscopic masses computed from two independent calibrations of $M_V$ -- one based on spectral classification and the other on FW3414 -- and combined with observed $T_{\rm eff}$ and $\log g$, against their directly determined counterparts. The agreement is overall satisfactory, with differences in $M_V$, radius, and luminosity generally within the expected uncertainties. This validates the use of both $M_V$ calibrations for deriving reliable estimates of radii and luminosities. In contrast, spectroscopic mass shows larger scatter and systematic deviations, especially above $40\,M_\odot$, suggesting that the mass calibration should be used cautiously and interpreted only in a statistical context.

The calibrations presented here, along with the accompanying tables that provide mean values and standard deviations for each SpT--LC bin, offer a robust framework for characterizing and calibrating O-type stars. All derived data products and the associated spectral library are made publicly available to support the community. These calibrations are particularly relevant for population or statistical studies where distances or full spectroscopic modeling are unavailable -- such as in extragalactic settings -- though we emphasize that they are optimized for Galactic stars with high-quality data. Applications to more distant, unresolved, or heavily reddened systems should be approached carefully, as extinction, metallicity, or classification uncertainties may introduce systematics beyond the scope of this study.

Our current analysis is limited to stars with reliable distances. The upcoming {\it Gaia} DR4 release would improve parallax estimates for many stars excluded here. Moreover, large-scale spectroscopic surveys such as WEAVE and 4MOST will greatly expand the number of O-type stars with high-resolution spectra and good parallaxes. 
In future work, we plan to combine the observed parameters of the \numCalib\ calibrator stars with the calibrated parameters derived for the remaining \numElim\ stars in our database. Studying how the full sample behaves with respect to expected empirical relations, and specially the properties of the fundamental relations, will allow us to refine our understanding of the main physical characteristics of stars across the O-star domain.

\section*{Data availability}
Table~\ref{tableValues} is only fully available in electronic form at the CDS via \url{https://cdsarc.cds.unistra.fr/viz-bin/cat/J/A+A/703/A175}.
The spectra and the parameters derived from them are available in the ICAOB database \url{https://research.iac.es/proyecto/iacob/iacobcat/}. The remaining parameters used in this work including: spectral classification, Gaia parallax and photometry, and Bailer-Jones distance estimates, are accessible online through the respective catalogs referenced in the text.

\begin{acknowledgements}

G.H. acknowledges support from the State Research Agency (AEI) of the Spanish Ministry of Science and Innovation (MICIN) and the European Regional Development Fund, FEDER under grants LOS MÚLTIPLES CANALES DE EVOLUCIÓN TEMPRANA DE LAS ESTRELLAS MASIVAS/ LA ESTRUCTURA DE LA VIA LACTEA DESVELADA POR SUS ESTRELLAS MASIVAS with reference PID2021-122397NB-C21 / PID2022-136640NB-C22 / 10.13039/501100011033. 

This project received the support from the “La Caixa” Foundation (ID 100010434) under the fellowship code LCF/BQ/PI23/11970035. 

The project leading to this application has received funding from European Commission (EC) under Project OCEANS - Overcoming challenges in the evolution and nature of massive stars, HORIZON-MSCA-2023-SE-01, No G.A 101183150
Funded by the European Union. Views and opinions expressed are however those of the author(s) only and do not necessarily reflect those of the European Union or the European Research Executive Agency (REA). Neither the European Union nor the granting authority can be held responsible for them. 

This work has made use of data from the European Space Agency (ESA) mission {\it Gaia} (\url{https://www.cosmos.esa.int/gaia}), processed by the {\it Gaia} Data Processing and Analysis Consortium (DPAC, \url{https://www.cosmos.esa.int/web/gaia/dpac/consortium}). Funding for the DPAC has been provided by national institutions, in particular the institutions participating in the {\it Gaia} Multilateral Agreement.

Based on observations made with the Nordic Optical Telescope, operated by NOTSA, and the Mercator Telescope, operated by the Flemish Community, both at the Observatorio del Roque de los Muchachos (La Palma, Spain) of the Instituto de Astrofísica de Canarias.  
Based on observations at the European Southern Observatory in programs 073.D-0609(A), 077.B-0348(A), 079.D-0564(A), 079.D-0564(C), 081.D-2008(A), 081.D-2008(B), 083.D-0589(A), 083.D-0589(B), 086.D-0997(A), 086.D-0997(B), 087.D-0946(A), 089.D-0975(A). 

The work reported on in this publication has been partially supported by COST Action CA18104: MW-Gaia.

\end{acknowledgements}

\bibliographystyle{aa}
\bibliography{sample}

\begin{appendix}

\section{Quality cuts for the calibrator sample}\label{AppCuts}

To ensure we use only stars with the most reliable fundamental parameters for calibration, we applied several quality control checks. The first filter targets stars with the most accurate distance measurements. The second filter focuses on photometric magnitudes, removing stars with uncertain or problematic values. Lastly, we examined extinction parameters (specifically Av), excluding stars with significant discrepancies with our reference study. This ensures a clean and dependable sample for calibration.

\subsection{Reliable Gaia distances}\label{AppCutsDist}

We applied a series of filters to retain stars with the most trustworthy distance measurements.

\begin{figure}
\includegraphics[width=0.5\textwidth]{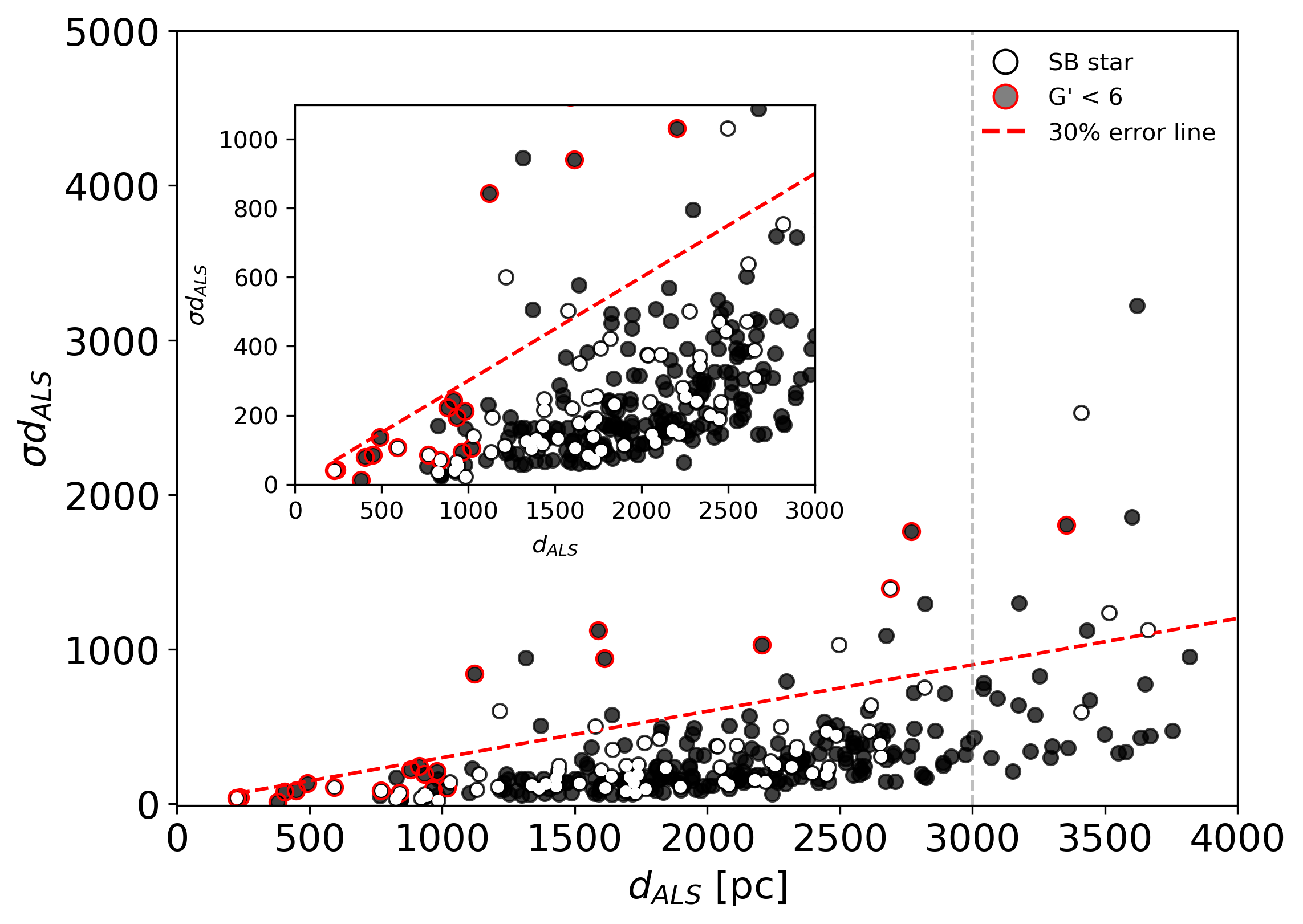}
\caption{ 
Relative distance uncertainty versus distance for the \numO\ stars in the working sample, based on ALS distances (see Sect. 5.2). Stars above the 30\% threshold or beyond 3000~pc were excluded. Red circles indicate stars discarded due to Gaia brightness issues.
}
\label{d_vs_ed_VH}
\end{figure}

First, it is known that Gaia encounters issues with extremely bright stars ($G_{corr}$ < 6, \citet{Martin-Fleitas2014,MaizApellaniz2017}), leading us to exclude 27 stars with $G_{corr}$ values below 6\footnote{Recent improvements in data processing allow for acceptable-quality astrometry down to $G \sim 3$ for some sources \citep{Fabricius2021}, although this requires a dedicated, star-by-star assessment.}. As is shown in Fig.~\ref{d_vs_ed_VH}, these stars are mostly within 1000 pc, but six fall between 1000 and 3000 pc. Notably, these stars also exhibit large relative distance errors, another exclusion criterion (discussed below).

\begin{figure}
\includegraphics[width=0.5\textwidth]{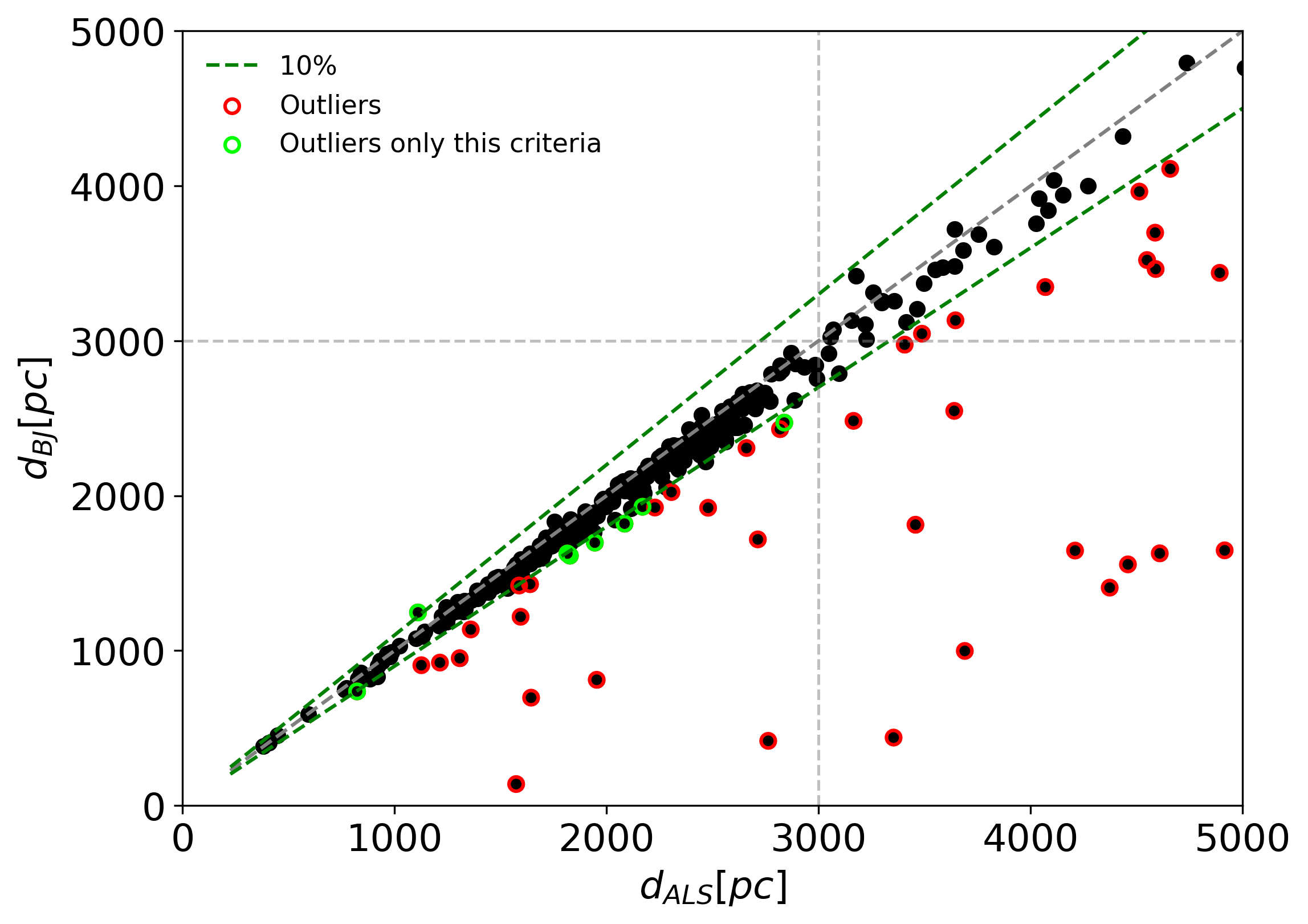}
\caption{Comparison between ALS and \cite{Bailer-Jones2021} distance estimates for 353 stars in the working sample. Stars with discrepancies larger than 10\% were excluded, and green points indicate stars rejected solely due to this criterion.
}
\label{dist_vs}
\end{figure}

The second filter excludes stars with ALS-calculated distances exceeding 3000 pc, removing 75 stars. ALS distances are refined through several steps, including zero-point corrections \citep{MaizApellaniz2021} and the use of a prior to account for the galactic disk, as massive stars are statistically clustered in this region \citep{Pantaleoni2025}. As is shown in Fig.~\ref{d_vs_ed_VH}, the majority of the sample is concentrated below 3000 pc. Beyond this distance, the sample size drops sharply, leading to less reliable distance estimates. This trend is further supported by comparisons with Bailer-Jones' results (Fig.~\ref{dist_vs}), discussed later.

The third filter removes stars with relative distance errors exceeding 30\%, eliminating 59 stars. However, Fig.~\ref{d_vs_ed_VH} reveals that only 9 stars were removed solely based on this criterion, as many had already been excluded due to brightness or distance.

Additionally, we excluded 6 bright stars for which ALS provided distances not derived from Gaia. To maintain uniformity, we opted to use only Gaia-based distances.

Lastly, we excluded 66 stars where the discrepancy between our distances and those from Bailer-Jones exceeded 10\%. We chose this conservative threshold to ensure the highest-quality stars for calibration. Figure~\ref{dist_vs} shows that only 8 stars were removed solely for this reason. The fact that most removed stars met multiple exclusion criteria supports the validity of their removal.

\subsection{Photometry and extinction cuts}\label{AppCutsPhot}

\begin{figure}
\includegraphics[width=0.5\textwidth]{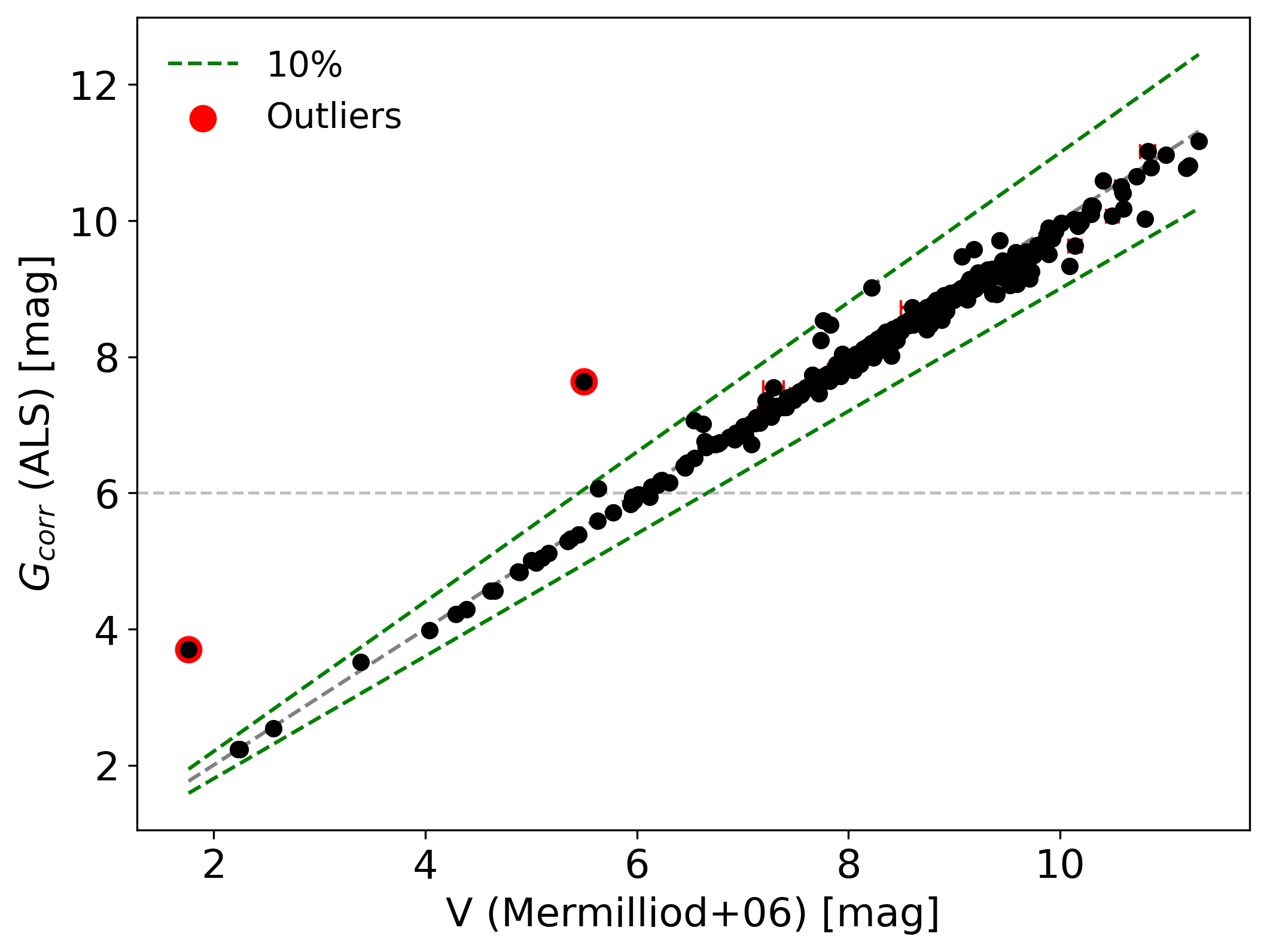}
\caption{Comparison between $G_{\rm corr}$ (\textit{Gaia}) and $V$ (Mermilliod) magnitudes. This relation was used to identify outliers with potentially unreliable Mermilliod photometry. 
}
\label{VvsGcorr}
\end{figure}

For photometry, we gathered data from five additional filters (BVJHK) alongside Gaia's $G_{corr}$. To validate Mermilliod's B and V photometry, we compared V to $G_{corr}$, as there is an expected direct correlation \citep{MaizApellaniz2018b}. Fig.~\ref{VvsGcorr} shows two clear outliers with differences greater than 10\%. These are HD~37743 and HD~113904~B. HD~37743 is a magnetic very bright star without ALS distance available, and HD113904B is a binary star likely blended in Mermilliod’s data but resolved by Gaia. These two stars were removed.

The JHK measurements from 2MASS show high quality overall. Out of \numO\ stars, only 23 lacked the AAA quality flag, and of those, just 3 had a lower U flag, indicating poor quality. HD~118198 AUU, HD~319703~B AUU, HD~34078 AAU. These three stars were removed from the calibrator sample.

\begin{figure}
\includegraphics[width=0.5\textwidth]{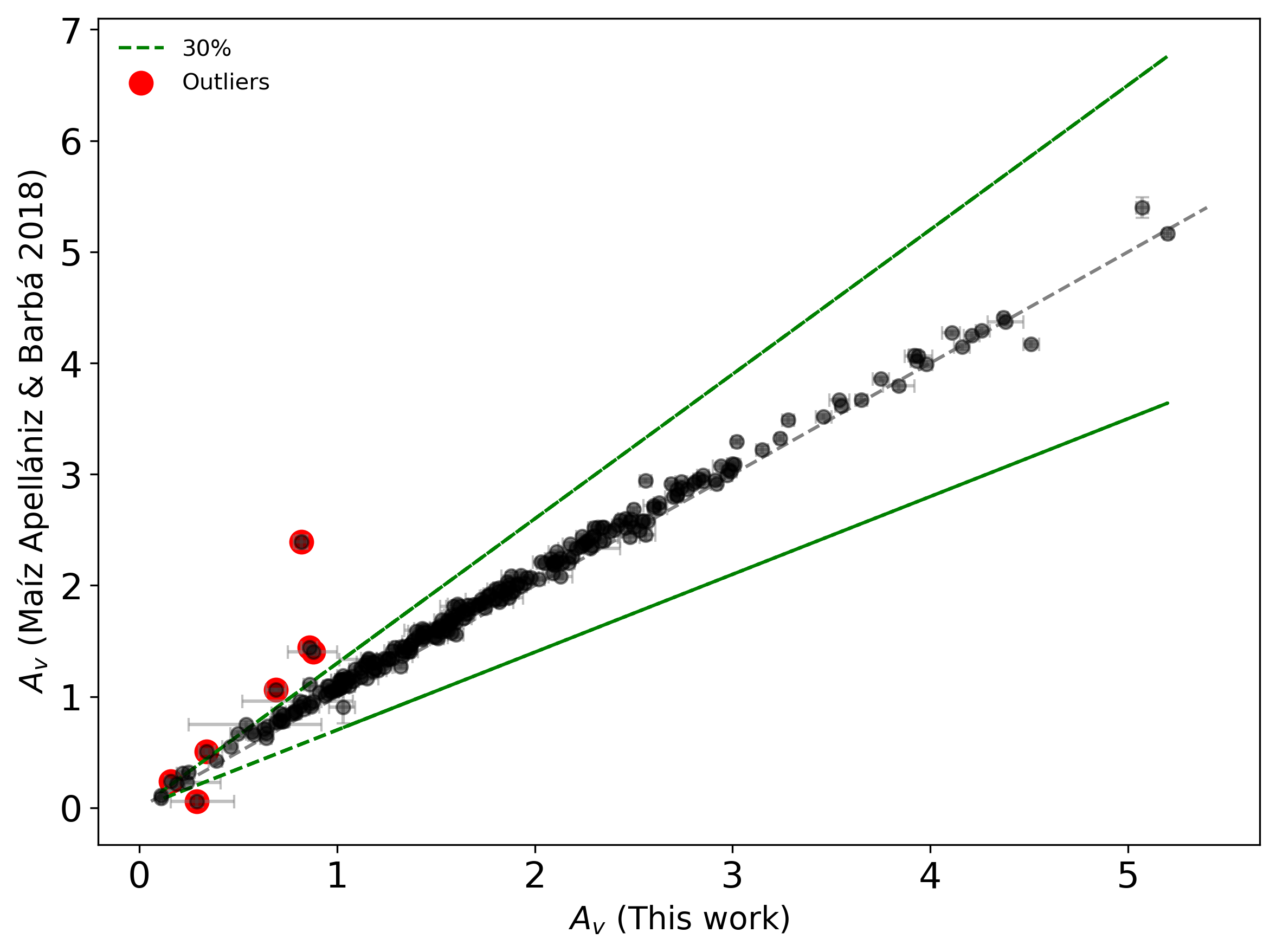}
\caption{Comparison between $A_V$ values derived in this work and those from \citet{MaizApellaniz2018} for the 289 stars in our working sample with available data in that reference. Stars with differences greater than 30\% were excluded from the calibrator sample. Only three were rejected solely based on this criterion.
}
\label{comp_Av}
\end{figure}

Since we calculated extinction values for the stars, we compared our results, specifically Av values, with those from \citep{MaizApellaniz2018}. As is shown in Fig.~\ref{comp_Av}, most of our values compare favorably, increasing our confidence in the extinction estimates. However, we removed seven stars that had discrepancies greater than 30\%, although only three were removed solely due to this criterion.

\subsection{Selection of the calibrator sample}\label{Selection}

A series of quality control checks were applied to ensure the selection of only the stars with the most reliable results, referred to as calibrators. These procedures are explained in detail above. Here, we summarize the selection criteria and report the final number of stars remaining after the cuts.

The first criterion was to select only \textit{Gaia} distances of the highest quality. To achieve this, we applied five specific filters. First, stars with $G_{corr}$ $<$ 6 were excluded due to known issues with \textit{Gaia}’s accuracy for bright stars (27 stars removed). 
Second, we limited the sample to stars with distances below 3000 pc, as beyond this limit the number of stars with reliable distance estimates decreases sharply, which affects the Galactic disk prior used in distance calculations (75 stars removed this way). 
Third, we required a relative error of less than 30\% in distance estimates to remove stars with clearly unreliable values (59 stars removed). Fourth, stars showing a discrepancy greater than 10\% between our calculated distance and the Bailer-Jones distance were excluded, as this indicates unreliable measurements (66 stars removed). Lastly, six bright stars were excluded because their distances were available from literature, but were not derived from \textit{Gaia} parallaxes.

Furthermore, two stars were removed based on photometric checks, where their V magnitude differed significantly from the $G_{corr}$ values used. We also excluded three stars whose 2MASS photometry had a U quality flag, indicating unreliable data. Finally, seven more stars were removed because their calculated Av values differed more than 30\% from those in our reference study, \cite{MaizApellaniz2018}. 

In the end, fundamental parameters were calculated for all \numO\ stars, with \numCalib\ classified as reliable and \numElim\ as unreliable. These distinctions are clearly indicated in the tables and figures.

\section{Calibration Formulas}

\begin{table}[t!]
\caption{Linear fits of the observed data presented in Fig.~\ref{Calibs_Tg_calibs}. \Teff\ (top) and \grav\ (bottom) as a function of SpTs for three LCs, with SPT the spectral type number of O-type stars.}
\label{Cab1}
        \begin{tabular}{r@{\hskip 0.01in}l}
                \Teff\ [kK] $=$ & $\left\{\begin{tabular}{@{\ }ll} -2.03  $\times$ SPT + 49.22 & (I, 36 stars) \\-1.88  $\times$ SPT + 49.42  & (III, 42 stars) \\ -1.75  $\times$ SPT + 50.01 & (V, 86 stars)
                \end{tabular}\right.$ \\
\\
        \grav\ [dex] $=$        & $\left\{\begin{tabular}{@{\ }ll} -0.10  $\times$ SPT + 4.15 & (I, 36 stars) \\-0.04  $\times$ SPT + 3.86  & (III, 42 stars) \\ { }0.012  $\times$ SPT + 3.81 & (V, 86 stars)
        \end{tabular}\right.$ \\
        \end{tabular}
\end{table}

\begin{table}[h!]
\caption{Linear fits of the observed data presented in Fig.~\ref{Calibs_Mv}. \Mv\ as a function of SpTs for three LCs, with SPT the spectral type number of O-type stars.}
\label{Cab2}
        \begin{tabular}{r@{\hskip 0.01in}l}
                \Mv\ [mag] $=$ & $\left\{\begin{tabular}{@{\ }ll} 0.039  $\times$ SPT - 6.26 & (I, 36 stars) \\0.288  $\times$ SPT - 7.370  & (III, 42 stars) \\ 0.240  $\times$ SPT - 6.202 & (V, 86 stars)
                \end{tabular}\right.$ \\
        \end{tabular}
\end{table}

\begin{table}[h!]
\caption{Linear fits of the observed data presented in Fig.~\ref{Calibs_RL_calibs}. Radii (top) and luminosity (bottom) as a function of SpTs for three LCs, with SPT the spectral type number of O-type stars.}
\label{Cab3}
        \begin{tabular}{r@{\hskip 0.01in}l}
                \(R\) [\(R_\odot\)] $=$ & $\left\{\begin{tabular}{@{\ }ll} 0.30  $\times$ SPT + 16.03 & (I, 36 stars) \\-1.17  $\times$ SPT + 21.39  & (III, 42 stars) \\ -0.69  $\times$ SPT + 13.52 & (V, 86 stars)
                \end{tabular}\right.$ \\
\\
        \Llum\ [dex] $=$        & $\left\{\begin{tabular}{@{\ }ll} -0.07  $\times$ SPT + 6.14 & (I, 36 stars) \\-0.18  $\times$ SPT + 6.69  & (III, 42 stars) \\ { }-0.15  $\times$ SPT + 6.18 & (V, 86 stars)
        \end{tabular}\right.$ \\
        \end{tabular}
\end{table}

\begin{table}[h!]
\caption{Parabolic fits of the observed data presented in Fig.~\ref{Calibs_RL_calibs}. \Msp\ as a function of SpTs for three LCs, with SPT the spectral type number of O-type stars.}
\label{Cab4}
        \begin{tabular}{r@{\hskip 0.01in}l}
                \Msp\ [\msol] $=$ & $\left\{\begin{tabular}{@{\ }ll} 1.20  $\times$ SPT$^2$ -23.00 $\times$ SPT + 133.44 & (I, 36 stars) \\2.11  $\times$ SPT$^2$ -39.42 $\times$ SPT + 196.57  & (III, 42 stars) \\ 0.45  $\times$ SPT$^2$ -9.65 $\times$ SPT + 67.49 & (V, 86 stars)
                \end{tabular}\right.$ \\
        \end{tabular}
\end{table}

\begin{table}[h!]
\caption{Linear fits of the observed data presented in Fig.~\ref{Calibs_BC}. \BCv\ as a function of SpTs for three LCs, with SPT the spectral type number of O-type stars.}
\label{CabBC}
        \begin{tabular}{r@{\hskip 0.01in}l}
                \BCv  $=$ & $\left\{\begin{tabular}{@{\ }ll} 0.174  $\times$ SPT - 4.54 & (I, 36 stars) \\0.170  $\times$ SPT - 4.62  & (III, 42 stars) \\ 0.140  $\times$ SPT - 4.50 & (V, 86 stars)
                \end{tabular}\right.$ \\
        \end{tabular}
\end{table}

\begin{table}[h!]
\caption{Linear fits of the observed data presented in Fig.~\ref{Calibs_FW3414_calibs}. \Mv\ (top) and \Lsp\ (bottom) as a function of FW3414 for three LCs, with FW344 of the H$\beta$ line in \AA. Additionally, parabolic fit of \Mv\ (middle) as a function of FW3414 independent of LC.}
\label{Cab5}
        \begin{tabular}{r@{\hskip 0.01in}l}
                \Mv\ [mag] $=$ & $\left\{\begin{tabular}{@{\ }ll} 0.24  $\times$ FW3414 - 6.77 & (I, 36 stars) \\0.56  $\times$ FW3414 - 7.77  & (III, 42 stars) \\ 0.32  $\times$ FW3414 - 6.34 & (V, 86 stars)
                \end{tabular}\right.$ \\
                \\
    \Mv\ [mag] $=$ & $\left\{\begin{tabular}{@{\ }ll} -0.055  $\times$ FW3414$^2$ +1.06 $\times$ FW3414 - 8.79 &  \\
                                                (All LC, 234 stars) &  \\
                \end{tabular}\right.$ \\ 
\\
        \Lsp\ $=$   & $\left\{\begin{tabular}{@{\ }ll} -0.023  $\times$ FW3414 + 4.22 & (I, 36 stars) \\-0.168  $\times$ FW3414 + 4.79  & (III, 42 stars) \\ { }-0.151  $\times$ FW3414 + 4.65 & (V, 86 stars)
        \end{tabular}\right.$ \\
        \end{tabular}
\end{table}

\section{Figures}

\begin{figure}
\includegraphics[width=0.5\textwidth]{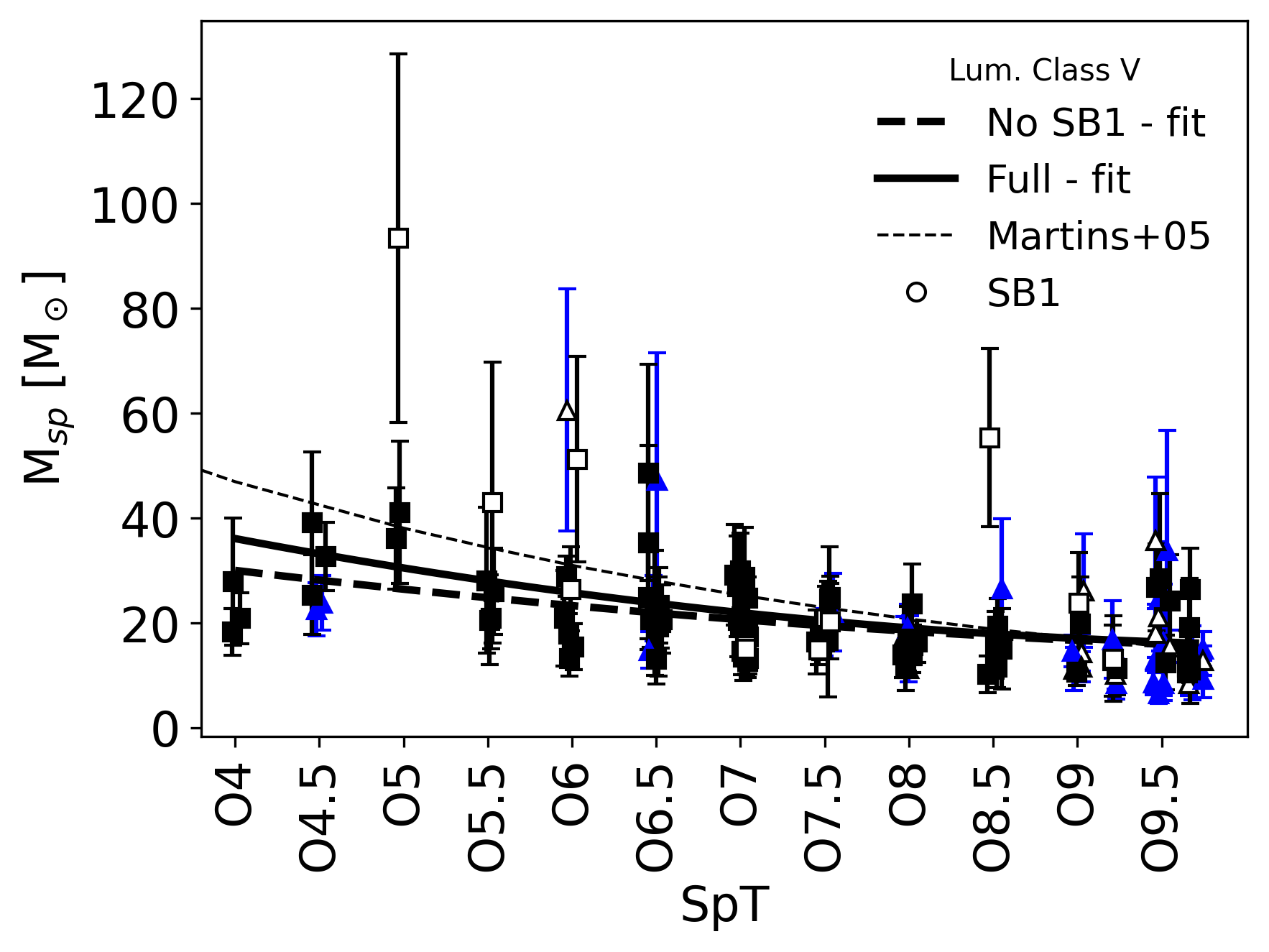}
\caption{Similar to Fig.~\ref{Calibs_RL_calibs}, panel bottom-right. Impact of SB1 stars (white symbols) on the spectroscopic mass calibration for LC~V stars. 
The solid black line shows the polynomial fit derived from the full calibrator sample, including SB1 systems. The dashed line corresponds to the fit obtained after removing all SB1 stars. }
\label{Msp_LCV_comp_SB1}
\end{figure}

\begin{figure}
\includegraphics[width=0.5\textwidth]{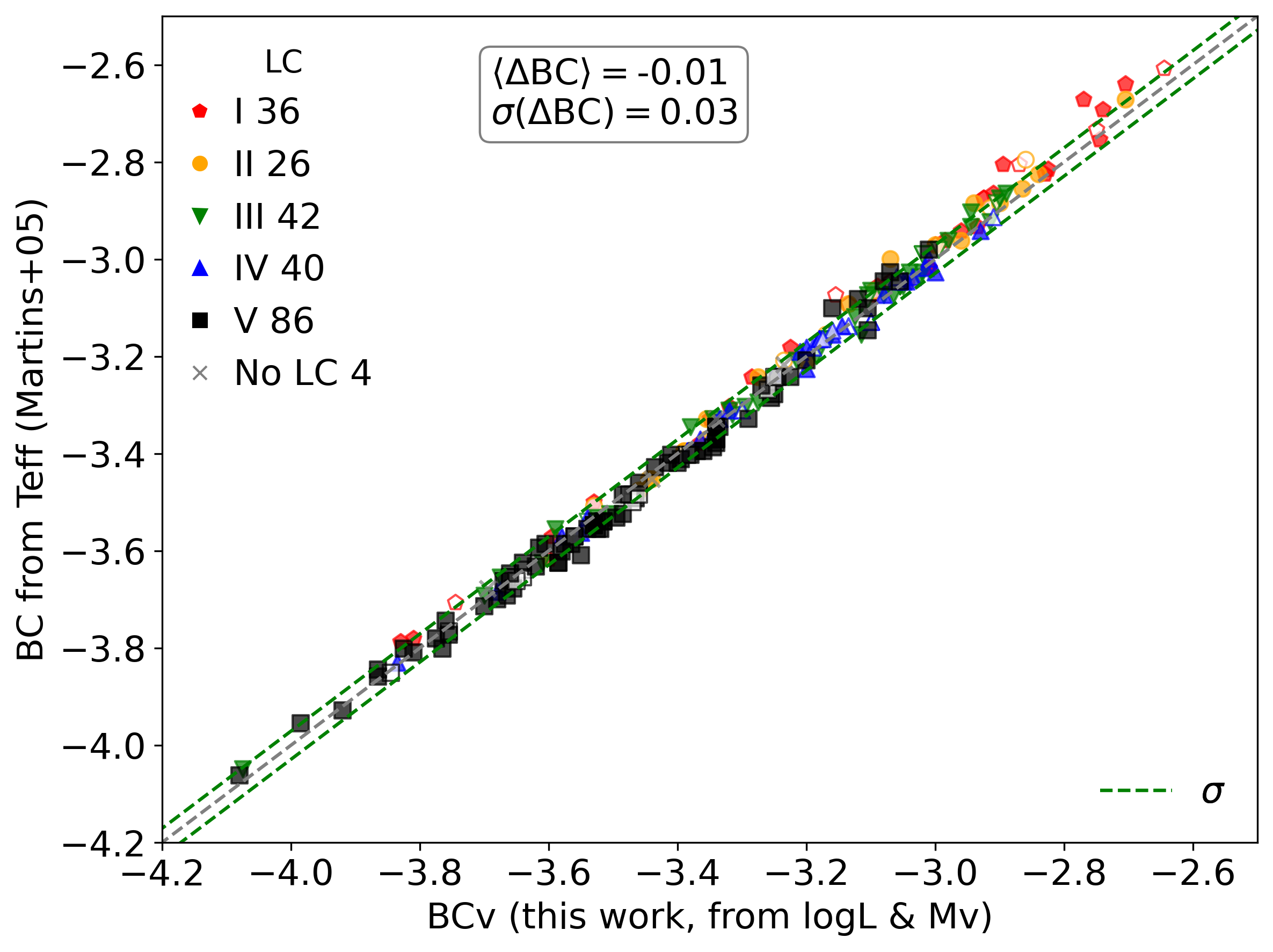}
\caption{Comparison between the empirical bolometric corrections (\BCv) derived in this work and those computed using the $T_{\rm eff}$–BC relation from M05. The dashed line marks the one-to-one relation, and $\pm$1 $\sigma$.}
\label{BCvsMartins}
\end{figure}

\section{Tables}

Tables~\ref{tableValues_LC_Teff} to \ref{tableValues_LC_Msp} summarize the results of our calibrations for each SpT and LC bin. For each parameter, we report the number of stars contributing to the bin, the mean and standard deviation of the observed values, the value obtained from our calibration, and, for comparison, the corresponding entry from the observational scale of M05.

\onecolumn
\begin{landscape}
        \pagestyle{empty}
		\fontsize{8}{8}\selectfont
\onecolumn

{
\setlength\LTleft{-7.5cm}  

\noindent\begin{longtable}{l@{\hskip 0.13in}l@{\hskip 0.13in}l@{\hskip 0.13in}c@{\hskip 0.13in}c@{\hskip 0.13in}c@{\hskip 0.13in}c@{\hskip 0.13in}c@{\hskip 0.13in}c@{\hskip 0.13in}c@{\hskip 0.13in}c@{\hskip 0.13in}c@{\hskip 0.13in}c@{\hskip 0.13in}c@{\hskip 0.13in}c@{\hskip 0.13in}c}
\captionsetup{justification=raggedright, singlelinecheck=false}
\caption{Spectroscopic and fundamental parameters for the calibrator sample curated in this work. }\\
\label{tableValues}\\ 
\hline \hline
\noalign{\smallskip}
         ID &              SpC & SB &    \#sp. &  \vsini\ &  \Teff &   \gravt  &   $G_{corr}$  &              \Av &          $d_{ALS}$ &               \Mv &              R &            \Llum &            \Msp &              \BCv &  FW3414 \\
          &               & status &    & [\kms]&    [kK] &    [dex]&    [mag] &               [mag]&           [pc]&               [mag]&               [R$_{\odot}$]&             [dex]&            [M$_{\odot}$]&              [mag] &  [\AA] \\         
\hline
\noalign{\smallskip}
\endfirsthead
\caption{continued.}\\
\hline
\hline
\noalign{\smallskip}
         ID &              SpC & SB &    \#sp. &  \vsini\ &  \Teff &   \gravt  &   $G_{corr}$  &              \Av &          $d_{ALS}$ &               \Mv &              R &            \Llum &            \Msp &              \BCv &  FW3414 \\
          &               & status &    & [\kms]&    [kK] &    [dex]&    [mag] &               [mag]&           [pc]&               [mag]&               [R$_{\odot}$]&             [dex]&            [M$_{\odot}$]&              [mag] &  [\AA] \\         
\hline
\noalign{\smallskip}
\endhead
\noalign{\smallskip}
\noalign{\smallskip}
\hline\hline
\noalign{\smallskip}
\multicolumn{16}{l}{\textbf{SB status:} \textbf{C}:Constant star, \textbf{LPV}:Line profile variability, \textbf{SB1}:Single-lined spectroscopic binary. Classifications based on fewer than three spectra (as indicated in the \#sp. column) should be treated with caution.}\\
\multicolumn{16}{l}{\parbox{23.7cm}{\textbf{Notes: }star ID; spectral classification (SpC) from the ALS catalog; spectroscopic binarity status (SB), following the methodology of \citet{Holgado2019} and \citet{MartinezSebastian2024}; number of spectra available; projected rotational velocity ($v \sin i$) derived with the {\sc iacob-broad} tool assuming a 10\% uncertainty; effective temperature ($T_{\rm eff}$) and rotationally corrected surface gravity ($\log g_{\rm true}$), obtained from the {\sc iacob-gbat/fastwind} analysis; corrected Gaia G-band magnitude ($G_{\rm corr}$) following \citet{MaizApellaniz2018b}, with average error of 0.003; visual extinction in V band; and distance from the ALS catalog ($d_{\rm ALS}$). The remaining parameters were computed following the methodology described in Sects.\ref{section3} and~\ref{section42}. \BCv\ refers to the bolometric correction derived from the stellar luminosity, and FW3414 represents the difference between the width of the H$\beta$ measured at three-quarters and one-quarter of its line depth, with global uncertainty of 0.1 \AA. The full table is available at the CDS.}}\\
\endfoot

   ALS12619 &        O7V((f))z &        C &         1 &        20 & 38.3 $\pm$ 0.8 & 4.08 $\pm$ 0.16 & 10.496 & 2.35 $\pm$ 0.08 & 2670 $\pm$ 140 & -3.92 $\pm$ 0.15 &  6.3 $\pm$ 0.4 & 4.89 $\pm$ 0.07 &  17.4 $\pm$ 6.4 & -3.56 $\pm$ 0.23 &        5.8 \\
    ALS4880 &         O6V((f)) &        C &         1 &        84 & 39.2 $\pm$ 1.2 & 3.91 $\pm$ 0.18 & 10.073 & 3.84 $\pm$ 0.08 & 1750 $\pm$ 110 & -4.58 $\pm$ 0.17 &  8.5 $\pm$ 0.7 & 5.19 $\pm$ 0.09 &  20.9 $\pm$ 9.1 & -3.66 $\pm$ 0.28 &        4.9 \\
    ALS5039 &       ON6V((f))z &        C &         7 &       123 & 41.2 $\pm$ 0.6 & 4.22 $\pm$ 0.04 & 10.804 & 3.98 $\pm$ 0.03 &  1955 $\pm$ 94 & -4.22 $\pm$ 0.11 &  6.9 $\pm$ 0.4 & 5.09 $\pm$ 0.05 &  28.9 $\pm$ 3.9 & -3.76 $\pm$ 0.17 &        5.4 \\
  BD+331025 &        O7.5V(n)z &        C &         1 &       177 & 38.6 $\pm$ 1.7 & 4.21 $\pm$ 0.29 & 10.209 &  1.8 $\pm$ 0.03 & 2590 $\pm$ 240 &  -3.56 $\pm$ 0.2 &  5.3 $\pm$ 0.5 & 4.74 $\pm$ 0.11 & 17.0 $\pm$ 11.0 & -3.55 $\pm$ 0.34 &        6.0 \\
  BD+364063 &          ON9.7Ib &      SB1 &         6 &       116 & 27.5 $\pm$ 1.1 & 3.03 $\pm$ 0.13 &  9.143 & 4.26 $\pm$ 0.03 &  1728 $\pm$ 72 &  -5.74 $\pm$ 0.1 & 18.6 $\pm$ 0.8 & 5.25 $\pm$ 0.08 &  13.8 $\pm$ 4.4 & -2.64 $\pm$ 0.22 &        2.0 \\
  BD+364145 &         O8.5V(n) &        C &         8 &       203 & 35.3 $\pm$ 1.3 & 3.74 $\pm$ 0.19 &  8.733 & 2.63 $\pm$ 0.03 &  1440 $\pm$ 65 &   -4.5 $\pm$ 0.1 &  8.8 $\pm$ 0.4 & 5.03 $\pm$ 0.07 &  15.8 $\pm$ 6.4 &  -3.34 $\pm$ 0.2 &        5.2 \\
   BD+56594 &             O7Vz &        C &         3 &        26 & 38.3 $\pm$ 0.7 & 3.98 $\pm$ 0.12 &  9.637 & 1.83 $\pm$ 0.04 & 2420 $\pm$ 160 & -3.96 $\pm$ 0.15 &  6.5 $\pm$ 0.5 & 4.91 $\pm$ 0.07 &  14.6 $\pm$ 4.5 & -3.58 $\pm$ 0.23 &        5.7 \\
   BD+60134 &    O5.5V(n)((f)) &        C &         4 &       249 & 41.3 $\pm$ 1.7 & 4.03 $\pm$ 0.19 & 10.402 & 2.72 $\pm$ 0.03 & 2710 $\pm$ 150 &  -4.3 $\pm$ 0.12 &  7.3 $\pm$ 0.4 & 5.14 $\pm$ 0.09 &  20.6 $\pm$ 8.6 & -3.81 $\pm$ 0.25 &        5.7 \\
   BD+60261 &  O7.5III(n)((f)) &        C &         3 &       161 & 35.0 $\pm$ 0.4 & 3.53 $\pm$ 0.05 &  8.585 &  1.8 $\pm$ 0.04 & 2570 $\pm$ 250 & -5.22 $\pm$ 0.21 & 12.4 $\pm$ 1.2 & 5.31 $\pm$ 0.09 &  19.0 $\pm$ 4.2 & -3.31 $\pm$ 0.31 &        5.0 \\
  BD+602635 &         O6V((f)) &        C &         5 &        50 & 39.7 $\pm$ 0.9 &  3.77 $\pm$ 0.1 & 10.018 & 2.29 $\pm$ 0.04 & 2820 $\pm$ 180 & -4.42 $\pm$ 0.13 &  7.8 $\pm$ 0.5 & 5.13 $\pm$ 0.07 &  13.3 $\pm$ 3.4 & -3.66 $\pm$ 0.22 &        3.9 \\
   BD+60499 &            O9.5V &        C &         1 &        17 & 34.6 $\pm$ 0.9 & 3.97 $\pm$ 0.18 & 10.135 & 2.48 $\pm$ 0.06 &  1976 $\pm$ 85 & -3.68 $\pm$ 0.12 &  6.0 $\pm$ 0.3 & 4.67 $\pm$ 0.06 &  12.4 $\pm$ 5.2 & -3.26 $\pm$ 0.19 &        6.6 \\
   \multicolumn{16}{c}{...}\\
    HD96622 &           O9.2IV &      SB1 &         4 &        39 & 32.8 $\pm$ 0.7 & 3.64 $\pm$ 0.11 &  8.828 & 1.43 $\pm$ 0.04 & 2180 $\pm$ 160 & -4.21 $\pm$ 0.15 &  8.0 $\pm$ 0.6 & 4.82 $\pm$ 0.07 &  10.3 $\pm$ 2.9 &  -3.1 $\pm$ 0.23 &        5.6 \\
    HD96638 &            O8III &        C &         1 &       249 & 33.7 $\pm$ 0.8 &  3.5 $\pm$ 0.08 &  8.463 & 1.73 $\pm$ 0.04 & 2350 $\pm$ 180 & -5.02 $\pm$ 0.17 & 11.5 $\pm$ 0.9 & 5.19 $\pm$ 0.08 &  15.5 $\pm$ 3.4 & -3.22 $\pm$ 0.26 &        5.1 \\
    HD96654 &            O9III &        C &         2 &       112 & 33.2 $\pm$ 0.4 & 3.57 $\pm$ 0.07 &  8.479 & 1.42 $\pm$ 0.04 & 2520 $\pm$ 290 & -4.87 $\pm$ 0.25 & 10.9 $\pm$ 1.3 &  5.11 $\pm$ 0.1 &  16.1 $\pm$ 4.3 & -3.16 $\pm$ 0.35 &        6.1 \\
    HD96715 &        O4V((f))z &        C &         2 &        59 & 45.0 $\pm$ 2.1 & 3.86 $\pm$ 0.16 &  8.188 & 1.33 $\pm$ 0.03 & 2590 $\pm$ 300 & -5.13 $\pm$ 0.25 & 10.2 $\pm$ 1.2 & 5.58 $\pm$ 0.13 & 28.0 $\pm$ 12.0 & -4.08 $\pm$ 0.41 &        4.1 \\
    HD96917 &       O8Ib(n)(f) &        C &         4 &       165 & 32.0 $\pm$ 0.2 & 3.25 $\pm$ 0.05 &  7.016 & 1.16 $\pm$ 0.04 & 2680 $\pm$ 470 &  -6.22 $\pm$ 0.4 & 21.1 $\pm$ 3.9 & 5.62 $\pm$ 0.16 & 29.0 $\pm$ 10.0 & -3.09 $\pm$ 0.57 &        3.3 \\
    HD96946 &       O6.5III(f) &        C &         5 &        71 & 38.8 $\pm$ 0.7 & 3.84 $\pm$ 0.11 &  8.384 & 1.58 $\pm$ 0.04 & 2660 $\pm$ 430 & -5.25 $\pm$ 0.35 & 11.7 $\pm$ 1.9 & 5.44 $\pm$ 0.14 & 35.0 $\pm$ 14.0 & -3.61 $\pm$ 0.49 &        4.3 \\
    HD97253 &         O5III(f) &        C &         4 &        70 & 39.2 $\pm$ 0.4 & 3.62 $\pm$ 0.05 &  7.034 & 1.41 $\pm$ 0.03 & 2420 $\pm$ 430 & -6.21 $\pm$ 0.37 & 18.4 $\pm$ 3.1 & 5.85 $\pm$ 0.15 & 52.0 $\pm$ 18.0 & -3.68 $\pm$ 0.53 &        3.5 \\
    HD97319 &      O7.5IV((f)) &        C &         1 &        51 & 35.6 $\pm$ 0.7 &  3.61 $\pm$ 0.1 &  8.475 & 1.59 $\pm$ 0.06 & 2700 $\pm$ 330 & -5.22 $\pm$ 0.27 & 12.1 $\pm$ 1.5 & 5.33 $\pm$ 0.11 &  22.0 $\pm$ 7.4 & -3.36 $\pm$ 0.39 &        4.4 \\
    HD97434 &  O7.5III(n)((f)) &        C &         3 &       217 & 34.9 $\pm$ 0.4 & 3.56 $\pm$ 0.04 &  8.036 & 1.43 $\pm$ 0.03 & 2480 $\pm$ 330 & -5.34 $\pm$ 0.28 & 13.1 $\pm$ 1.7 & 5.36 $\pm$ 0.11 &  22.5 $\pm$ 5.6 & -3.32 $\pm$ 0.39 &        5.2 \\
    HD97848 &              O8V &        C &         2 &        41 & 35.6 $\pm$ 0.7 & 3.66 $\pm$ 0.08 &  8.618 &  1.0 $\pm$ 0.04 & 2660 $\pm$ 480 &  -4.42 $\pm$ 0.4 &  8.4 $\pm$ 1.5 &  5.0 $\pm$ 0.16 &  11.7 $\pm$ 4.6 & -3.34 $\pm$ 0.57 &        5.1 \\
    HD99897 &      O6.5IV((f)) &        C &         1 &        50 & 37.6 $\pm$ 0.6 & 3.56 $\pm$ 0.06 &  8.261 & 1.56 $\pm$ 0.04 & 2260 $\pm$ 220 & -4.97 $\pm$ 0.22 & 10.5 $\pm$ 1.1 &  5.3 $\pm$ 0.09 &  14.8 $\pm$ 3.5 & -3.54 $\pm$ 0.31 &        3.8 \\
     \end{longtable}
}

\end{landscape}

{\fontsize{8}{8}\selectfont
\setlength\LTleft{-0.8cm}


\noindent\begin{longtable}{crrrrrrrrrrrrrrr}
\caption{\Teff~ calibrations}
\label{tableValues_LC_Teff} \\
\hline \hline
\noalign{\smallskip}
& \multicolumn{5}{c}{x\,=\,\Teff\ (kK)\ -- LC V} & \multicolumn{5}{c}{x\,=\,\Teff\ (kK)\ -- LC III} & \multicolumn{5}{c}{x\,=\,\Teff\ (kK)\ -- LC I} \\
\cmidrule(lr){2-6} \cmidrule(lr){7-11} \cmidrule(lr){12-16}
 ST &  \# &  $\langle x \rangle$ &  $\sigma(x)$ &  x (fit) &  x (M05) &  \# &  $\langle x \rangle$ &  $\sigma(x)$ &  x (fit) &  x (M05) &  \# &  $\langle x \rangle$ &  $\sigma(x)$ &  x (fit) &  x (M05)  \\
\hline
\noalign{\smallskip}
\endfirsthead
\caption{continued.}\\
\hline
\hline
\noalign{\smallskip}
& \multicolumn{5}{c}{x\,=\,\Teff\ (kK)\ -- LC V} & \multicolumn{5}{c}{x\,=\,\Teff\ (kK)\ -- LC III} & \multicolumn{5}{c}{x\,=\,\Teff\ (kK)\ -- LC I} \\
\cmidrule(lr){2-6} \cmidrule(lr){7-11} \cmidrule(lr){12-16}
 ST &  \# &  $\langle x \rangle$ &  $\sigma(x)$ &  x (fit) &  x (M05) &  \# &  $\langle x \rangle$ &  $\sigma(x)$ &  x (fit) &  x (M05) &  \# &  $\langle x \rangle$ &  $\sigma(x)$ &  x (fit) &  x (M05)  \\
\hline
\noalign{\smallskip}
\endhead
\noalign{\smallskip}
\noalign{\smallskip}
\hline\hline
\noalign{\smallskip}
\endfoot
       4.0 &     3 &       42.16 &       0.60 &      43.01 &      43.41 &     1 &       44.80 &          - &      41.91 &      41.48 &     3 &       40.73 &       0.43 &      41.06 &      40.70 \\
       4.5 &     3 &       41.83 &       0.48 &      42.14 &          - &     0 &           - &          - &      40.96 &          - &     0 &           - &          - &      40.04 &          - \\
       5.0 &     3 &       41.20 &       0.69 &      41.27 &      41.54 &     2 &       38.55 &       0.65 &      40.02 &      39.50 &     1 &       38.10 &          - &      39.02 &      38.52 \\
       5.5 &     5 &       39.28 &       0.60 &      40.39 &      40.06 &     0 &           - &          - &      39.07 &      38.00 &     0 &           - &          - &      38.00 &      37.07 \\
       6.0 &     7 &       40.03 &       0.82 &      39.52 &      38.15 &     1 &       38.70 &          - &      38.12 &      36.67 &     1 &       38.70 &          - &      36.99 &      35.74 \\
       6.5 &    10 &       38.34 &       0.60 &      38.65 &      36.82 &     4 &       37.96 &       0.56 &      37.18 &      35.64 &     1 &       37.20 &          - &      35.97 &      34.65 \\
       7.0 &    17 &       37.61 &       0.99 &      37.77 &      35.53 &     0 &           - &          - &      36.23 &      34.63 &     2 &       35.36 &       0.30 &      34.95 &      33.32 \\
       7.5 &     8 &       37.12 &       0.69 &      36.90 &      34.41 &     4 &       35.08 &       0.14 &      35.29 &      33.48 &     4 &       33.35 &       1.04 &      33.94 &      31.91 \\
       8.0 &     7 &       36.19 &       0.60 &      36.03 &      33.38 &     1 &       33.70 &          - &      34.34 &      32.57 &     2 &       32.01 &       0.07 &      32.92 &      31.00 \\
       8.5 &     6 &       36.08 &       0.91 &      35.15 &      32.52 &     3 &       33.24 &       1.32 &      33.39 &      31.68 &     1 &       31.90 &          - &      31.90 &      30.50 \\
       9.0 &     3 &       34.59 &       0.94 &      34.28 &      31.52 &     9 &       32.51 &       0.56 &      32.45 &      30.73 &     3 &       31.92 &       0.30 &      30.89 &      29.56 \\
       9.2 &     3 &       34.23 &       0.15 &      33.93 &          - &     3 &       31.83 &       0.36 &      32.07 &          - &     4 &       31.40 &       0.65 &      30.48 &          - \\
       9.5 &     4 &       34.60 &       0.69 &      33.40 &      30.48 &     4 &       31.21 &       0.59 &      31.50 &      30.23 &     1 &       30.80 &          - &      29.87 &      28.43 \\
       9.7 &     7 &       31.77 &       0.46 &      33.05 &          - &    10 &       31.20 &       0.80 &      31.12 &          - &    13 &       29.37 &       1.00 &      29.46 &          - \\
    \end{longtable}   


\noindent\begin{longtable}{crrrrrrrrrrrrrrr}
\caption{\grav~ calibrations}
\label{tableValues_LC_grav}\\ 
\hline \hline
\noalign{\smallskip}
& \multicolumn{5}{c}{x\,=\,\grav\ (dex)\ -- LC V} & \multicolumn{5}{c}{x\,=\,\grav\ (dex)\ -- LC III} & \multicolumn{5}{c}{x\,=\,\grav\ (dex)\ -- LC I} \\
\cmidrule(lr){2-6} \cmidrule(lr){7-11} \cmidrule(lr){12-16}
 ST &  \# &  $\langle x \rangle$ &  $\sigma(x)$ &  x (fit) &  x (M05) &  \# &  $\langle x \rangle$ &  $\sigma(x)$ &  x (fit) &  x (M05) &  \# &  $\langle x \rangle$ &  $\sigma(x)$ &  x (fit) &  x (M05)  \\
\hline
\noalign{\smallskip}
\endfirsthead
\caption{continued.}\\
\hline
\hline
\noalign{\smallskip}
& \multicolumn{5}{c}{x\,=\,\grav\ (dex)\ -- LC V} & \multicolumn{5}{c}{x\,=\,\grav\ (dex)\ -- LC III} & \multicolumn{5}{c}{x\,=\,\grav\ (dex)\ -- LC I} \\
\cmidrule(lr){2-6} \cmidrule(lr){7-11} \cmidrule(lr){12-16}
 ST &  \# &  $\langle x \rangle$ &  $\sigma(x)$ &  x (fit) &  x (M05) &  \# &  $\langle x \rangle$ &  $\sigma(x)$ &  x (fit) &  x (M05) &  \# &  $\langle x \rangle$ &  $\sigma(x)$ &  x (fit) &  x (M05)  \\
\hline
\noalign{\smallskip}
\endhead
\noalign{\smallskip}
\noalign{\smallskip}
\hline\hline
\noalign{\smallskip}
\endfoot
       4.0 &     3 &        3.71 &       0.05 &       3.86 &       3.92 &     1 &        3.85 &          - &       3.71 &       3.73 &     3 &        3.67 &       0.02 &       3.73 &       3.65 \\
       4.5 &     3 &        3.83 &       0.05 &       3.87 &          - &     0 &           - &          - &       3.69 &          - &     0 &           - &          - &       3.67 &          - \\
       5.0 &     3 &        3.91 &       0.14 &       3.87 &       3.92 &     2 &        3.52 &       0.06 &       3.67 &       3.69 &     1 &        3.62 &          - &       3.62 &       3.57 \\
       5.5 &     5 &        3.74 &       0.07 &       3.88 &       3.92 &     0 &           - &          - &       3.65 &       3.67 &     0 &           - &          - &       3.57 &       3.52 \\
       6.0 &     7 &        4.05 &       0.18 &       3.88 &       3.92 &     1 &        3.78 &          - &       3.64 &       3.65 &     1 &        3.67 &          - &       3.52 &       3.48 \\
       6.5 &    10 &        3.79 &       0.11 &       3.89 &       3.92 &     4 &        3.63 &       0.12 &       3.62 &       3.63 &     1 &        3.52 &          - &       3.46 &       3.44 \\
       7.0 &    17 &        3.76 &       0.14 &       3.90 &       3.92 &     0 &           - &          - &       3.60 &       3.61 &     2 &        3.46 &       0.09 &       3.41 &       3.40 \\
       7.5 &     8 &        3.80 &       0.13 &       3.90 &       3.92 &     4 &        3.46 &       0.04 &       3.58 &       3.59 &     4 &        3.25 &       0.10 &       3.36 &       3.36 \\
       8.0 &     7 &        3.78 &       0.11 &       3.91 &       3.92 &     1 &        3.38 &          - &       3.56 &       3.57 &     2 &        3.20 &          - &       3.31 &       3.32 \\
       8.5 &     6 &        4.05 &       0.21 &       3.91 &       3.92 &     3 &        3.46 &       0.14 &       3.54 &       3.55 &     1 &        3.24 &          - &       3.25 &       3.28 \\
       9.0 &     3 &        3.91 &       0.10 &       3.92 &       3.92 &     9 &        3.41 &       0.08 &       3.52 &       3.53 &     3 &        3.28 &       0.04 &       3.20 &       3.23 \\
       9.2 &     3 &        3.92 &       0.02 &       3.92 &          - &     3 &        3.34 &       0.06 &       3.52 &          - &     4 &        3.26 &       0.05 &       3.18 &          - \\
       9.5 &     4 &        4.03 &       0.03 &       3.93 &       3.92 &     4 &        3.44 &       0.11 &       3.51 &       3.51 &     1 &        3.13 &          - &       3.15 &       3.19 \\
       9.7 &     7 &        3.83 &       0.11 &       3.93 &          - &    10 &        3.43 &       0.16 &       3.50 &          - &    13 &        3.05 &       0.09 &       3.13 &          - \\
    \end{longtable}   


\noindent\begin{longtable}{crrrrrrrrrrrrrrr}
\caption{\Mv~ calibrations}\\
\label{tableValues_LC_Mv}\\ 
\hline \hline
\noalign{\smallskip}
& \multicolumn{5}{c}{x\,=\,\Mv\ (mag)\ -- LC V} & \multicolumn{5}{c}{x\,=\,\Mv\ (mag)\ -- LC III} & \multicolumn{5}{c}{x\,=\,\Mv\ (mag)\ -- LC I} \\
\cmidrule(lr){2-6} \cmidrule(lr){7-11} \cmidrule(lr){12-16}
 ST &  \# &  $\langle x \rangle$ &  $\sigma(x)$ &  x (fit) &  x (M05) &  \# &  $\langle x \rangle$ &  $\sigma(x)$ &  x (fit) &  x (M05) &  \# &  $\langle x \rangle$ &  $\sigma(x)$ &  x (fit) &  x (M05)  \\
\hline
\noalign{\smallskip}
\endfirsthead
\caption{continued.}\\
\hline
\hline
\noalign{\smallskip}
& \multicolumn{5}{c}{x\,=\,\Mv\ (mag)\ -- LC V} & \multicolumn{5}{c}{x\,=\,\Mv\ (mag)\ -- LC III} & \multicolumn{5}{c}{x\,=\,\Mv\ (mag)\ -- LC I} \\
\cmidrule(lr){2-6} \cmidrule(lr){7-11} \cmidrule(lr){12-16}
 ST &  \# &  $\langle x \rangle$ &  $\sigma(x)$ &  x (fit) &  x (M05) &  \# &  $\langle x \rangle$ &  $\sigma(x)$ &  x (fit) &  x (M05) &  \# &  $\langle x \rangle$ &  $\sigma(x)$ &  x (fit) &  x (M05)  \\
\hline
\noalign{\smallskip}
\endhead
\noalign{\smallskip}
\noalign{\smallskip}
\hline\hline
\noalign{\smallskip}
\endfoot
       4.0 &     3 &    -4.913 &     0.12 &   -5.243 &    -5.50 &     1 &    -6.410 &        - &   -6.206 &    -5.98 &     3 &    -5.994 &     0.42 &   -6.094 &    -6.34 \\
       4.5 &     3 &    -5.041 &     0.32 &   -5.123 &        - &     0 &         - &        - &   -6.063 &        - &     0 &         - &        - &   -6.075 &        - \\
       5.0 &     3 &    -5.126 &     0.16 &   -5.003 &    -5.21 &     2 &    -5.974 &     0.27 &   -5.919 &    -5.84 &     1 &    -6.290 &        - &   -6.057 &    -6.33 \\
       5.5 &     5 &    -4.922 &     0.25 &   -4.883 &    -5.07 &     0 &         - &        - &   -5.776 &    -5.76 &     0 &         - &        - &   -6.038 &    -6.33 \\
       6.0 &     7 &    -4.493 &     0.32 &   -4.763 &    -4.92 &     1 &    -5.510 &        - &   -5.633 &    -5.69 &     1 &    -5.660 &        - &   -6.020 &    -6.32 \\
       6.5 &    10 &    -4.620 &     0.31 &   -4.643 &    -4.77 &     4 &    -5.093 &     0.25 &   -5.490 &    -5.62 &     1 &    -5.330 &        - &   -6.001 &    -6.31 \\
       7.0 &    17 &    -4.545 &     0.43 &   -4.523 &    -4.63 &     0 &         - &        - &   -5.347 &    -5.54 &     2 &    -5.736 &     0.33 &   -5.982 &    -6.31 \\
       7.5 &     8 &    -4.473 &     0.51 &   -4.403 &    -4.48 &     4 &    -5.210 &     0.14 &   -5.204 &    -5.47 &     4 &    -6.073 &     0.22 &   -5.964 &    -6.30 \\
       8.0 &     7 &    -4.339 &     0.28 &   -4.283 &    -4.34 &     1 &    -5.020 &        - &   -5.061 &    -5.40 &     2 &    -6.111 &     0.04 &   -5.945 &    -6.30 \\
       8.5 &     6 &    -4.341 &     0.23 &   -4.163 &    -4.19 &     3 &    -4.926 &     0.64 &   -4.918 &    -5.32 &     1 &    -5.940 &        - &   -5.927 &    -6.29 \\
       9.0 &     3 &    -3.989 &     0.10 &   -4.043 &    -4.05 &     9 &    -4.990 &     0.22 &   -4.775 &    -5.25 &     3 &    -5.903 &     0.20 &   -5.908 &    -6.29 \\
       9.2 &     3 &    -3.753 &     0.02 &   -3.995 &        - &     3 &    -5.010 &     0.13 &   -4.718 &        - &     4 &    -5.920 &     0.43 &   -5.901 &        - \\
       9.5 &     4 &    -4.093 &     0.29 &   -3.923 &    -3.90 &     4 &    -4.122 &     0.75 &   -4.632 &    -5.18 &     1 &    -5.850 &        - &   -5.890 &    -6.28 \\
       9.7 &     7 &    -3.697 &     0.25 &   -3.875 &        - &    10 &    -4.501 &     0.30 &   -4.574 &        - &    13 &    -5.972 &     0.53 &   -5.882 &        - \\
    \end{longtable}   


\noindent\begin{longtable}{crrrrrrrrrrrrrrr}
\caption{R calibrations}\\
\label{tableValues_LC_R}\\ 
\hline \hline
\noalign{\smallskip}
& \multicolumn{5}{c}{x\,=\,$R$ ($R_{\odot}$) -- LC V} & \multicolumn{5}{c}{x\,=\,$R$ ($R_{\odot}$) -- LC III} & \multicolumn{5}{c}{x\,=\,$R$ ($R_{\odot}$) -- LC I} \\
\cmidrule(lr){2-6} \cmidrule(lr){7-11} \cmidrule(lr){12-16}
 ST &  \# &  $\langle x \rangle$ &  $\sigma(x)$ &  x (fit) &  x (M05) &  \# &  $\langle x \rangle$ &  $\sigma(x)$ &  x (fit) &  x (M05) &  \# &  $\langle x \rangle$ &  $\sigma(x)$ &  x (fit) &  x (M05)  \\
\hline
\noalign{\smallskip}
\endfirsthead
\caption{continued.}\\
\hline
\hline
\noalign{\smallskip}
& \multicolumn{5}{c}{x\,=\,$R$ ($R_{\odot}$) -- LC V} & \multicolumn{5}{c}{x\,=\,$R$ ($R_{\odot}$) -- LC III} & \multicolumn{5}{c}{x\,=\,$R$ ($R_{\odot}$) -- LC I} \\
\cmidrule(lr){2-6} \cmidrule(lr){7-11} \cmidrule(lr){12-16}
 ST &  \# &  $\langle x \rangle$ &  $\sigma(x)$ &  x (fit) &  x (M05) &  \# &  $\langle x \rangle$ &  $\sigma(x)$ &  x (fit) &  x (M05) &  \# &  $\langle x \rangle$ &  $\sigma(x)$ &  x (fit) &  x (M05)  \\
\hline
\noalign{\smallskip}
\endhead
\noalign{\smallskip}
\noalign{\smallskip}
\hline\hline
\noalign{\smallskip}
\endfoot
       4.0 &     3 &     10.0 &    0.45 &    11.0 &   12.31 &     1 &     18.0 &       - &    17.0 &   15.83 &     3 &     16.0 &    2.79 &    17.0 &   18.91 \\
       4.5 &     3 &     10.0 &    1.55 &    10.0 &       - &     0 &        - &       - &    16.0 &       - &     0 &        - &       - &    17.0 &       - \\
       5.0 &     3 &     11.0 &    0.73 &    10.0 &   11.08 &     2 &     16.0 &    1.89 &    16.0 &   15.26 &     1 &     19.0 &       - &    18.0 &   19.48 \\
       5.5 &     5 &     10.0 &    1.15 &    10.0 &   10.61 &     0 &        - &       - &    15.0 &   15.13 &     0 &        - &       - &    18.0 &   19.92 \\
       6.0 &     7 &      8.0 &    1.17 &     9.0 &   10.23 &     1 &     13.0 &       - &    14.0 &   14.97 &     1 &     14.0 &       - &    18.0 &   20.33 \\
       6.5 &    10 &      9.0 &    1.22 &     9.0 &    9.79 &     4 &     11.0 &    1.08 &    14.0 &   14.74 &     1 &     13.0 &       - &    18.0 &   20.68 \\
       7.0 &    17 &      8.0 &    1.65 &     9.0 &    9.37 &     0 &        - &       - &    13.0 &   14.51 &     2 &     15.0 &    2.09 &    18.0 &   21.14 \\
       7.5 &     8 &      8.0 &    1.94 &     8.0 &    8.94 &     4 &     12.0 &    0.83 &    13.0 &   14.34 &     4 &     19.0 &    2.03 &    18.0 &   21.69 \\
       8.0 &     7 &      8.0 &    1.02 &     8.0 &    8.52 &     1 &     12.0 &       - &    12.0 &   14.11 &     2 &     20.0 &    0.34 &    18.0 &   22.03 \\
       8.5 &     6 &      8.0 &    1.02 &     8.0 &    8.11 &     3 &     10.0 &    2.59 &    11.0 &   13.88 &     1 &     18.0 &       - &    19.0 &   22.20 \\
       9.0 &     3 &      7.0 &    0.33 &     7.0 &    7.73 &     9 &     11.0 &    1.32 &    11.0 &   13.69 &     3 &     18.0 &    1.61 &    19.0 &   22.60 \\
       9.2 &     3 &      6.0 &    0.07 &     7.0 &       - &     3 &     12.0 &    0.81 &    11.0 &       - &     4 &     18.0 &    3.31 &    19.0 &       - \\
       9.5 &     4 &      7.0 &    1.11 &     7.0 &    7.39 &     4 &      7.0 &    2.09 &    10.0 &   13.37 &     1 &     18.0 &       - &    19.0 &   23.11 \\
       9.7 &     7 &      6.0 &    0.88 &     7.0 &       - &    10 &      9.0 &    1.28 &    10.0 &       - &    13 &     19.0 &    3.43 &    19.0 &       - \\
    \end{longtable}   


\noindent\begin{longtable}{crrrrrrrrrrrrrrr}
\caption{\Llum~ calibrations}\\
\label{tableValues_LC_Llum}\\ 
\hline \hline
\noalign{\smallskip}
& \multicolumn{5}{c}{x\,=\,\Llum\ -- LC V} & \multicolumn{5}{c}{x\,=\,\Llum\ -- LC III} & \multicolumn{5}{c}{x\,=\,\Llum\ -- LC I} \\
\cmidrule(lr){2-6} \cmidrule(lr){7-11} \cmidrule(lr){12-16}
 ST &  \# &  $\langle x \rangle$ &  $\sigma(x)$ &  x (fit) &  x (M05) &  \# &  $\langle x \rangle$ &  $\sigma(x)$ &  x (fit) &  x (M05) &  \# &  $\langle x \rangle$ &  $\sigma(x)$ &  x (fit) &  x (M05)  \\
\hline
\noalign{\smallskip}
\endfirsthead
\caption{continued.}\\
\hline
\hline
\noalign{\smallskip}
& \multicolumn{5}{c}{x\,=\,\Llum\ -- LC V} & \multicolumn{5}{c}{x\,=\,\Llum\ -- LC III} & \multicolumn{5}{c}{x\,=\,\Llum\ -- LC I} \\
\cmidrule(lr){2-6} \cmidrule(lr){7-11} \cmidrule(lr){12-16}
 ST &  \# &  $\langle x \rangle$ &  $\sigma(x)$ &  x (fit) &  x (M05) &  \# &  $\langle x \rangle$ &  $\sigma(x)$ &  x (fit) &  x (M05) &  \# &  $\langle x \rangle$ &  $\sigma(x)$ &  x (fit) &  x (M05)  \\
\hline
\noalign{\smallskip}
\endhead
\noalign{\smallskip}
\noalign{\smallskip}
\hline\hline
\noalign{\smallskip}
\endfoot
       4.0 &     3 &        5.48 &       0.05 &       5.57 &       5.68 &     1 &        6.09 &          - &       5.96 &       5.82 &     3 &        5.85 &       0.19 &       5.85 &       5.94 \\
       4.5 &     3 &        5.49 &       0.11 &       5.50 &          - &     0 &           - &          - &       5.87 &          - &     0 &           - &          - &       5.82 &          - \\
       5.0 &     3 &        5.49 &       0.07 &       5.42 &       5.51 &     2 &        5.74 &       0.12 &       5.78 &       5.70 &     1 &        5.85 &          - &       5.78 &       5.87 \\
       5.5 &     5 &        5.30 &       0.11 &       5.34 &       5.41 &     0 &           - &          - &       5.69 &       5.63 &     0 &           - &          - &       5.74 &       5.82 \\
       6.0 &     7 &        5.17 &       0.12 &       5.27 &       5.30 &     1 &        5.58 &          - &       5.59 &       5.56 &     1 &        5.60 &          - &       5.71 &       5.78 \\
       6.5 &    10 &        5.19 &       0.15 &       5.19 &       5.20 &     4 &        5.36 &       0.11 &       5.50 &       5.49 &     1 &        5.44 &          - &       5.67 &       5.74 \\
       7.0 &    17 &        5.09 &       0.16 &       5.12 &       5.10 &     0 &           - &          - &       5.41 &       5.43 &     2 &        5.54 &       0.14 &       5.63 &       5.69 \\
       7.5 &     8 &        5.08 &       0.17 &       5.04 &       5.00 &     4 &        5.32 &       0.05 &       5.32 &       5.36 &     4 &        5.64 &       0.08 &       5.60 &       5.64 \\
       8.0 &     7 &        4.98 &       0.11 &       4.96 &       4.90 &     1 &        5.19 &          - &       5.23 &       5.30 &     2 &        5.60 &       0.01 &       5.56 &       5.60 \\
       8.5 &     6 &        4.94 &       0.12 &       4.89 &       4.82 &     3 &        5.18 &       0.23 &       5.14 &       5.24 &     1 &        5.50 &          - &       5.53 &       5.58 \\
       9.0 &     3 &        4.80 &       0.08 &       4.81 &       4.72 &     9 &        5.14 &       0.08 &       5.05 &       5.17 &     3 &        5.48 &       0.06 &       5.49 &       5.54 \\
       9.2 &     3 &        4.69 &       0.01 &       4.78 &          - &     3 &        5.11 &       0.04 &       5.01 &          - &     4 &        5.46 &       0.16 &       5.47 &          - \\
       9.5 &     4 &        4.82 &       0.12 &       4.74 &       4.62 &     4 &        4.78 &       0.31 &       4.95 &       5.12 &     1 &        5.42 &          - &       5.45 &       5.49 \\
       9.7 &     7 &        4.62 &       0.09 &       4.71 &          - &    10 &        4.90 &       0.12 &       4.92 &          - &    13 &        5.43 &       0.22 &       5.44 &          - \\
    \end{longtable}   


\noindent\begin{longtable}{crrrrrrrrrrrrrrr}
\caption{\Msp~ calibrations}\\
\label{tableValues_LC_Msp}\\ 
\hline \hline
\noalign{\smallskip}
& \multicolumn{5}{c}{x\,=\,$M_{\rm sp}$ ($M_{\odot}$) -- LC V} & \multicolumn{5}{c}{x\,=\,$M_{\rm sp}$ ($M_{\odot}$) -- LC III} & \multicolumn{5}{c}{x\,=\,$M_{\rm sp}$ ($M_{\odot}$) -- LC I} \\
\cmidrule(lr){2-6} \cmidrule(lr){7-11} \cmidrule(lr){12-16}
 ST &  \# &  $\langle x \rangle$ &  $\sigma(x)$ &  x (fit) &  x (M05) &  \# &  $\langle x \rangle$ &  $\sigma(x)$ &  x (fit) &  x (M05) &  \# &  $\langle x \rangle$ &  $\sigma(x)$ &  x (fit) &  x (M05)  \\
\hline
\noalign{\smallskip}
\endfirsthead
\caption{continued.}\\
\hline
\hline
\noalign{\smallskip}
& \multicolumn{5}{c}{x\,=\,$M_{\rm sp}$ ($M_{\odot}$) -- LC V} & \multicolumn{5}{c}{x\,=\,$M_{\rm sp}$ ($M_{\odot}$) -- LC III} & \multicolumn{5}{c}{x\,=\,$M_{\rm sp}$ ($M_{\odot}$) -- LC I} \\
\cmidrule(lr){2-6} \cmidrule(lr){7-11} \cmidrule(lr){12-16}
 ST &  \# &  $\langle x \rangle$ &  $\sigma(x)$ &  x (fit) &  x (M05) &  \# &  $\langle x \rangle$ &  $\sigma(x)$ &  x (fit) &  x (M05) &  \# &  $\langle x \rangle$ &  $\sigma(x)$ &  x (fit) &  x (M05)  \\
\hline
\noalign{\smallskip}
\endhead
\noalign{\smallskip}
\noalign{\smallskip}
\hline\hline
\noalign{\smallskip}
\endfoot
       4.0 &     3 &       20.0 &       2.0 &      36.0 &     46.16 &     1 &       88.0 &         - &      73.0 &     48.80 &     3 &       51.0 &      16.0 &      61.0 &     58.03 \\
       4.5 &     3 &       31.0 &       5.0 &      33.0 &         - &     0 &          - &         - &      62.0 &         - &     0 &          - &         - &      54.0 &         - \\
       5.0 &     3 &       40.0 &      12.0 &      30.0 &     37.28 &     2 &       30.0 &      11.0 &      52.0 &     41.48 &     1 &       58.0 &         - &      48.0 &     50.87 \\
       5.5 &     5 &       23.0 &       4.0 &      28.0 &     34.17 &     0 &          - &         - &      44.0 &     38.92 &     0 &          - &         - &      43.0 &     48.29 \\
       6.0 &     7 &       19.0 &       7.0 &      26.0 &     31.73 &     1 &       40.0 &         - &      36.0 &     36.38 &     1 &       35.0 &         - &      38.0 &     45.78 \\
       6.5 &    10 &       20.0 &       5.0 &      24.0 &     29.02 &     4 &       17.0 &       5.0 &      30.0 &     33.68 &     1 &       23.0 &         - &      34.0 &     43.10 \\
       7.0 &    17 &       17.0 &       5.0 &      22.0 &     26.52 &     0 &          - &         - &      24.0 &     31.17 &     2 &       21.0 &      10.0 &      31.0 &     40.91 \\
       7.5 &     8 &       18.0 &       4.0 &      20.0 &     24.15 &     4 &       16.0 &       4.0 &      20.0 &     29.06 &     4 &       25.0 &       6.0 &      28.0 &     39.17 \\
       8.0 &     7 &       15.0 &       2.0 &      19.0 &     21.95 &     1 &       15.0 &         - &      17.0 &     26.89 &     2 &       25.0 &       3.0 &      26.0 &     36.77 \\
       8.5 &     6 &       14.0 &       6.0 &      18.0 &     19.82 &     3 &       14.0 &       4.0 &      14.0 &     24.84 &     1 &       23.0 &         - &      24.0 &     33.90 \\
       9.0 &     3 &       12.0 &       4.0 &      17.0 &     18.03 &     9 &       13.0 &       3.0 &      13.0 &     23.07 &     3 &       23.0 &       2.0 &      23.0 &     31.95 \\
       9.2 &     3 &       12.0 &       1.0 &      17.0 &         - &     3 &       12.0 &       3.0 &      13.0 &         - &     4 &       19.0 &       6.0 &      23.0 &         - \\
       9.5 &     4 &       19.0 &       7.0 &      16.0 &     16.46 &     4 &        7.0 &       3.0 &      13.0 &     21.04 &     1 &       17.0 &         - &      23.0 &     30.41 \\
       9.7 &     7 &       12.0 &       2.0 &      16.0 &         - &    10 &       11.0 &       2.0 &      13.0 &         - &    13 &       15.0 &       6.0 &      23.0 &         - \\
    \end{longtable}   
}

\end{appendix}

	\end{document}